\documentclass[10pt]{iopart}

\makeatletter
\def\blfootnote{\gdef\@thefnmark{}\@footnotetext}
\makeatother

\expandafter\let\csname equation*\endcsname\relax
\expandafter\let\csname endequation*\endcsname\relax
\usepackage{amsmath}
\usepackage{amssymb}
\usepackage{amsfonts}
\usepackage[ruled,vlined,linesnumbered]{algorithm2e}
\usepackage{graphicx}
\usepackage{textcomp}
\usepackage{url}
\usepackage{float}
\usepackage{mathrsfs}

\usepackage{xcolor}
\definecolor{codegreen}{rgb}{0,0.6,0}
\definecolor{codegray}{rgb}{0.5,0.5,0.5}
\definecolor{codepurple}{rgb}{0.58,0,0.82}
\definecolor{backcolour}{rgb}{0.95,0.95,0.92}

\usepackage{listings}
\lstdefinestyle{mystyle}{
    backgroundcolor=\color{backcolour},   
    commentstyle=\color{codegreen},
    keywordstyle=\color{magenta},
    numberstyle=\tiny\color{codegray},
    stringstyle=\color{codepurple},
    basicstyle=\ttfamily,
    breakatwhitespace=false,         
    breaklines=true,                 
    captionpos=b,                    
    keepspaces=true,                 
    numbers=left,                    
    numbersep=5pt,                  
    showspaces=false,                
    showstringspaces=false,
    showtabs=false,                  
    tabsize=2
}
\usepackage{tikz}
\tikzstyle{arrow}=[->] 
\usetikzlibrary{calc}
\usetikzlibrary{bayesnet}
\usetikzlibrary{shapes.geometric}
\usetikzlibrary{shadows} 
\pgfdeclarelayer{fg}    
\pgfdeclarelayer{mid}    
\pgfdeclarelayer{bg}    
\pgfsetlayers{bg,mid,main,fg}  

\newcommand{\shortminus}{%
    \scalebox{0.75}[1.0]{\ensuremath{-}}
}

\usepackage{mathtools}
\usepackage{multirow}
\usepackage{hyperref}
\hypersetup{%
  colorlinks=false,
  linkbordercolor=red,
  pdfborderstyle={/S/U/W 1}
}

\begin{document}
\title[A Multi-label Approach to EMG-based Gesture Recognition]{A Multi-label Classification Approach to Increase Expressivity of EMG-based Gesture Recognition}
\author{
Niklas Smedemark-Margulies$^{\dagger,1}$
Yunus Bicer$^{\dagger,2}$,
Elifnur Sunger$^{2}$,
Stephanie Naufel$^{3}$,
Tales Imbiriba$^{2}$,
Eugene Tunik$^{4}$,
Deniz Erdo{\u{g}}mu{\c{s}}$^{2}$,
Mathew Yarossi$^{2,4}$
}
\address{$^{\dagger}$Equal Contribution}
\address{$^1$Khoury College of Computer Sciences, Northeastern University, Boston, MA, USA}
\address{$^2$Department of Electrical and Computer Engineering, Northeastern University, Boston, MA, USA}
\address{$^3$Meta Reality Labs Research, Menlo Park, CA}
\address{$^4$Department of Physical Therapy, Movement, and Rehabilitation Sciences, Northeastern University, Boston, MA, USA}
\ead{m.yarossi@northeastern.edu}

\begin{abstract}
\textit{Objective}.
The objective of the study is to efficiently increase the expressivity of surface electromyography-based (sEMG) gesture recognition systems. 
\textit{Approach}.
We use a problem transformation approach, in which actions were subset into two biomechanically independent components - a set of wrist directions and a set of finger modifiers. 
To maintain fast calibration time, we train models for each component using only individual gestures, and extrapolate to the full product space of combination gestures by generating synthetic data.
We collected a supervised dataset with high-confidence ground truth labels in which subjects performed combination gestures while holding a joystick, and conducted experiments to analyze the impact of model architectures, classifier algorithms, and synthetic data generation strategies on the performance of the proposed approach.
\textit{Main Results}.
We found that a problem transformation approach using a parallel model architecture in combination with a non-linear classifier, along with restricted synthetic data generation, shows promise in increasing the expressivity of sEMG-based gestures with a short calibration time.
\textit{Significance}.
sEMG-based gesture recognition has applications in human-computer interaction, virtual reality, and the control of robotic and prosthetic devices. 
Existing approaches require exhaustive model calibration.
The proposed approach increases expressivity without requiring users to demonstrate all combination gesture classes. 
Our results may be extended to larger gesture vocabularies and more complicated model architectures.
\end{abstract}

\vspace{2pc}

\maketitle
\ioptwocol

\section{Introduction}\label{sec:Introduction}
Surface electromyography (sEMG) provides a convenient sensor modality for human-computer interaction (HCI) applications~\cite{yang2021dynamic}. In the past two decades, research efforts have sought to translate the electrical activity associated with muscle contraction into control commands for general use computing, prosthetic control, and motor rehabilitation~\cite{xiong2021deep,qi2019intelligent}. sEMG-based gesture recognition describes the task of classifying hand gestures from an sEMG signal. To date, nearly all research efforts in this task have concentrated on the classification of single gestures. The expressivity of the gesture set can be greatly increased by allowing the combination of gestures; for example, combining a wrist movement to indicate a cursor direction, with a finger movement to indicate a mouse ``click.'' Combining gestures in this way results in a multi-label classification problem.

Algorithms for multi-label classification can be grouped broadly into \textit{algorithm adaptation} methods and \textit{problem transformation} methods~\cite{tsoumakas2006review}. The key difference is that algorithm adaptation focuses on designing a training objective to specifically address the multi-label problem, whereas problem transformation focuses on ways of combining multiple instances of existing single-label classifiers. The choice of algorithm adaptation or problem transformation is highly dependent upon the details of the classification task. The research focused on gesture compositions via multi-label classification is emerging, but to date remains relatively limited. 

High accuracy in classifying movement along multiple axes with application to prosthetic control has been achieved using fully supervised data that included all possible movement combinations~\cite{young2012classification, hahne2015concurrent, olsson2021learning}.
However, the requirement that all possible combinations of gestures be included in the training dataset limits the creation of expressive combinations of gestures, due to the time and effort needed to collect such exhaustive data. Therefore, an ideal combination gesture classification scheme would use only single gestures as training data. Few attempts have been made to classify combinations of gestures from single gestures. Several algorithm adaptation based approaches have utilized methods in which all classes that exceed a probability threshold are combined to form a composite gesture~\cite{olsson2019extraction, bjorklund2018investigating}. These approaches are not well suited for mutually exclusive combinations of gestures, as they may predict gesture combinations that are not biomechanically possible (i.e. simultaneous finger extension and flexion). 

This issue can be avoided by delineating labels into mutually exclusive groups and training a multi-label classifier. Problem transformation approaches utilizing separate classifiers for gesture sets are potentially well-suited to the multi-label classification of mutually exclusive gestures. To our knowledge, only a single study has investigated multi-label classification using a problem transformation approach. In that study, separate classifiers for flexion and extension of each digit were employed to construct multi-digit movements~\cite{mendes2022multi}. While novel, the utility of this approach given the large number of degrees of freedom of the hand is limited. 

In this work, we construct a vocabulary of gestures with two-part labels consisting of a direction component derived from wrist movement and a modifier component derived from finger movement. 
We then apply a problem transformation approach - in particular, we require that each prediction consists of exactly one direction and one modifier component. To limit calibration time we restrict training data to consist of only single direction and single modifier gestures. To accommodate this label structure, we use a model architecture in which an incoming gesture is simultaneously classified by two models; one model estimates the direction component and another model estimates the modifier component. Critically, we explore both a parallel and hierarchical implementation of the model architecture. 

We hypothesized that using a standard supervised learning approach with this model architecture and single gesture calibration data will unavoidably result in poor predictive accuracy for combination gestures.  We, therefore, explore a strategy to address this challenge using synthetic data augmentation; we collect a training set with only real single gestures and extrapolate to the set of combination gestures. In particular, given examples of two real single gestures, such as \lstinline{Up} and \lstinline{Pinch}, we can construct a synthetic combination \lstinline{Up, Pinch} gesture by blending their feature vectors.

We collect wrist worn sEMG data from healthy participants following instructions given via a custom user interface. Critically, we utilize a set of movements completed using a joystick to obtain ground truth labels for all actions and eliminate task error as a source of label noise. We perform three computational experiments in which we examine the model architecture and classification approach (Experiment 1), the selection of subsets of synthetic gesture combinations (Experiment 2), and the augmentation of single gestures to resolve class imbalances created by the creation of synthetic data (Experiment 3). We report results with respect to a lower bound model trained only on single gestures, and an upper bound model trained on both single gestures and real combination gestures. Through these experiments, we demonstrate the feasibility of using the problem transformation approach to multi-label classification of disjoint gesture sets from single label gesture training data with synthetic combinations. 

\section{Methods}
\subsection{Subjects}
A total of $11$ individuals ($6$ male / $5$ female, $25.18\pm 3.86$ years of age) participated in the study. All participants were right-handed, as confirmed via self-report. Prior to participating, subjects confirmed that they did not have any muscular, orthopedic, or neurological health issues that could affect their ability to perform the experiment. 

\subsection{Gesture Vocabulary}
\label{sec:vocab}
We defined an action space in which subjects may perform one of four possible direction gestures (\lstinline{Up}, \lstinline{Down}, \lstinline{Left}, \lstinline{Right}), one of two possible modifier gestures (\lstinline{Pinch}, \lstinline{Thumb}), and a rest gesture.
We explicitly generate expressivity by allowing combinations of one direction and one modifier, such that the total set of possible actions is $15$  ($4$ direction only, $2$ modifier only, $8$ combinations, and rest). We refer to combination gestures as ``doubles.''

\subsection{Sensors and Measurement}
\begin{figure}[!ht]
    \centering    
    \includegraphics[width=0.4\textwidth]{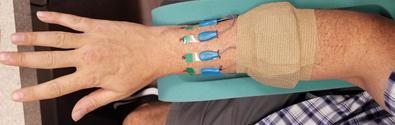}
    \includegraphics[width=0.4\textwidth]{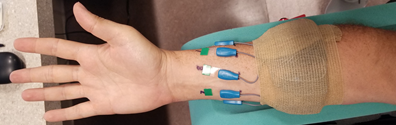}
    \caption{sEMG Recording. Electrodes were placed on the mid-forearm of the subject starting from mid-line on the dorsal aspect and continuing towards the thumb.}
    \label{fig:electrodes}
\end{figure}

Subjects were seated comfortably in front of an LCD with arms supported on arm rest, and the right hand positioned to a joystick. Surface electromyography  (sEMG, Trigno, Delsys Inc., 99.99\% Ag electrodes, $1926$ Hz sampling frequency, common mode rejection ratio: $>80$ dB, built-in 20\textendash450 Hz bandpass filter) was recorded from eight electrodes attached to the right forearm with adhesive tape. The eight electrodes were positioned with equidistant spacing around the circumference of the forearm at a four-finger-width distance from the ulnar styloid (the subject's left hand was wrapped around the right forearm at the ulnar styloid to determine the electrode placement); the first electrode was placed mid-line on the dorsal aspect of the forearm mid-line between the ulnar and radial styloid (see Figure~\ref{fig:electrodes}). 

Ground-truth labels were collected by simultaneously recording from the joystick (Logitech Extreme 3D Pro). The pitch axis (continuous) was used for Up (radial deviation) and Down (ulnar deviation) gestures, while the yaw axis (continuous) was used for Left (wrist flexion) and Right (wrist extension) gestures. We selected the direction with the greatest activation at any one time; joystick movement had to pass a threshold of $0.25$ on any given axis to be considered active. The trigger button (index finger flexion) was used for \lstinline{Pinch} gestures and the side thumb (thumb adduction) button for \lstinline{Thumb} gestures.
\begin{figure}[!ht]
    \centering    
    \includegraphics[width=0.485\textwidth]{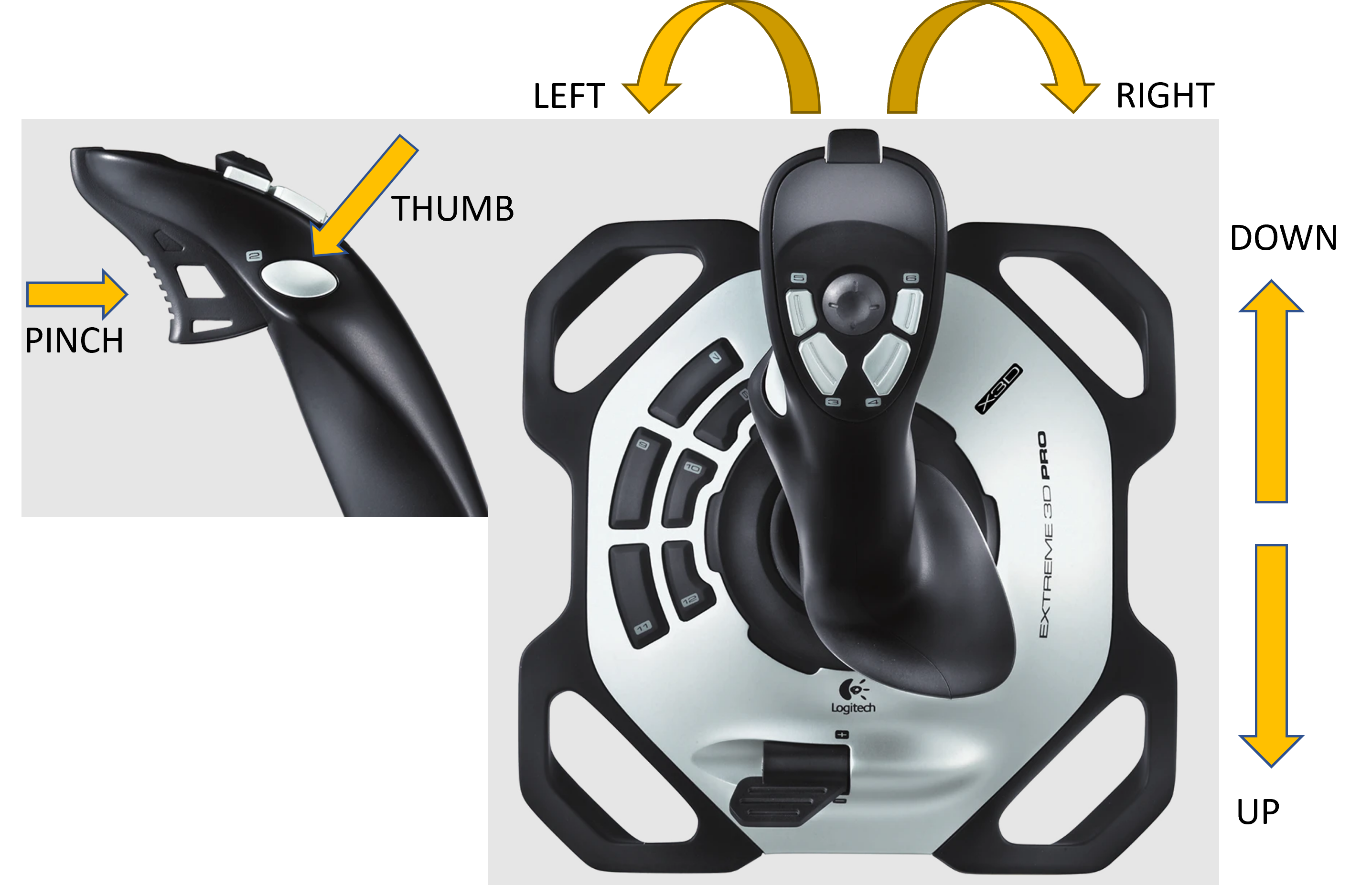}
    \caption{Ground-truth Movement Labels. Subjects manipulated a Logitech Extreme 3D Pro Joystick during all gestures, providing high-confidence ground truth labels. Direction movement was measured using pitch and yaw axes. Modifier movements used trigger and thumb buttons.}
    \label{fig:controller}
\end{figure}
To reduce the system jitter, we smoothed the output of the joystick using an exponential moving average with a momentum of $0.5$.

\subsection{Feature Extraction}
We extracted features from raw sEMG data using a sliding window. We chose a window size of $250$ms with $50$ms step size. From each of the $8$ sensor channels of raw sEMG, we computed the Root-Mean-Square (RMS) and Median Power Frequency after Fourier transform.
Given a data vector $x$ containing $T$ samples, the RMS is defined as:
\begin{align}
    \textrm{RMS}(x) = \sqrt{ \frac{1}{T}\sum_{i=1}^{T} x_{i}^2}.
\end{align}
The Median Power Frequency is defined as the frequency value $f_{\textsc{med}}$ that divides the integral of the Power Spectral Density (PSD) into two regions of equal area~\cite{hermens1992median}:
\begin{align}
    \int_{0}^{f_{\textsc{med}}} \!\textrm{PSD}(f) df = 
    \int_{f_{\textsc{med}}}^{\infty} \!\textrm{PSD}(f) df = 
    \frac{1}{2} \int_{0}^{\infty} \!\textrm{PSD}(f) df.
\end{align}

\subsection{Experimental Task}
\label{sec:Experimental_Task}
\begin{figure}[!ht]
    \centering
    \includegraphics[width=0.485\textwidth]{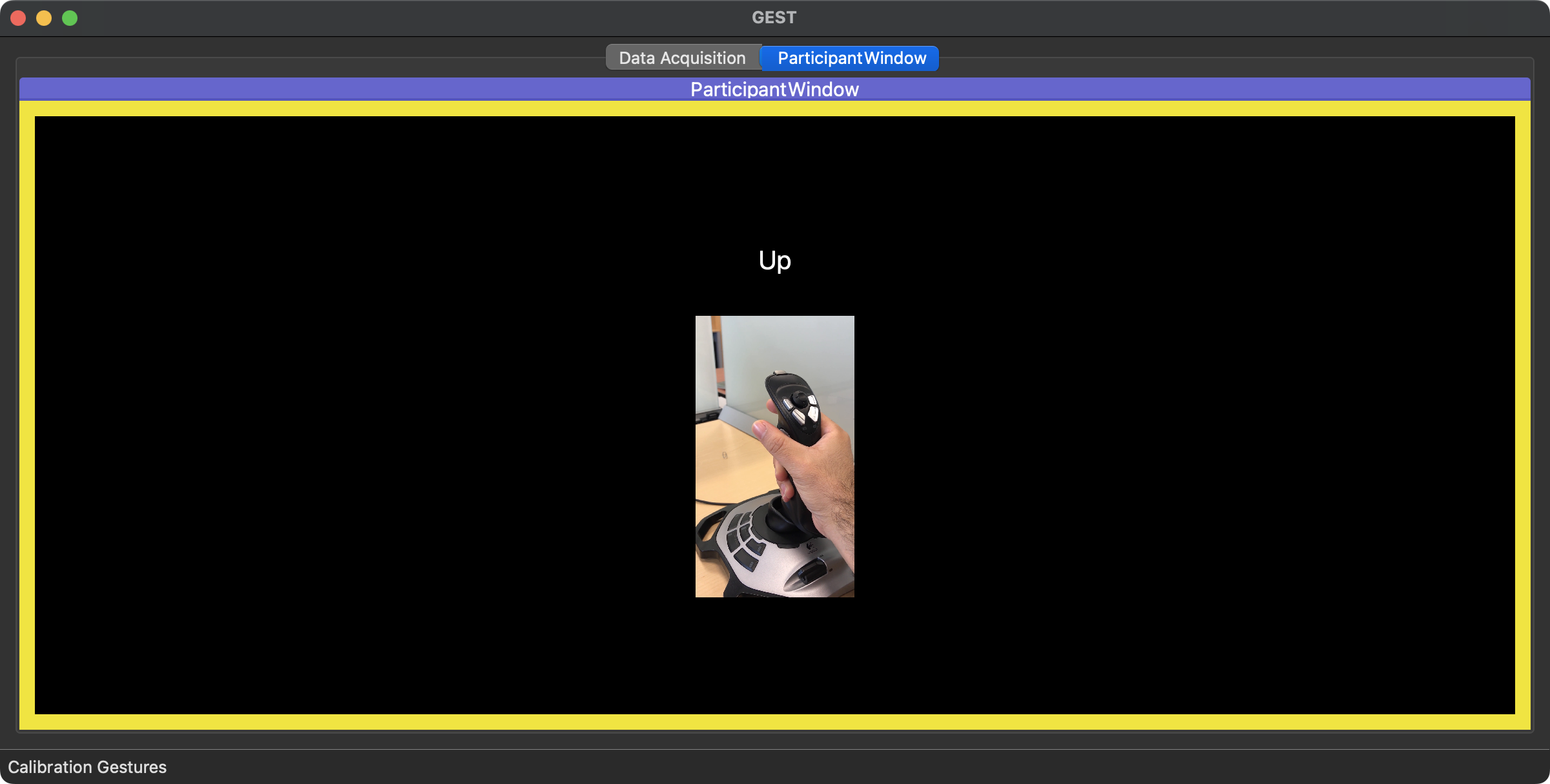}    
    \vspace{1pt}
    \includegraphics[width=0.485\textwidth]{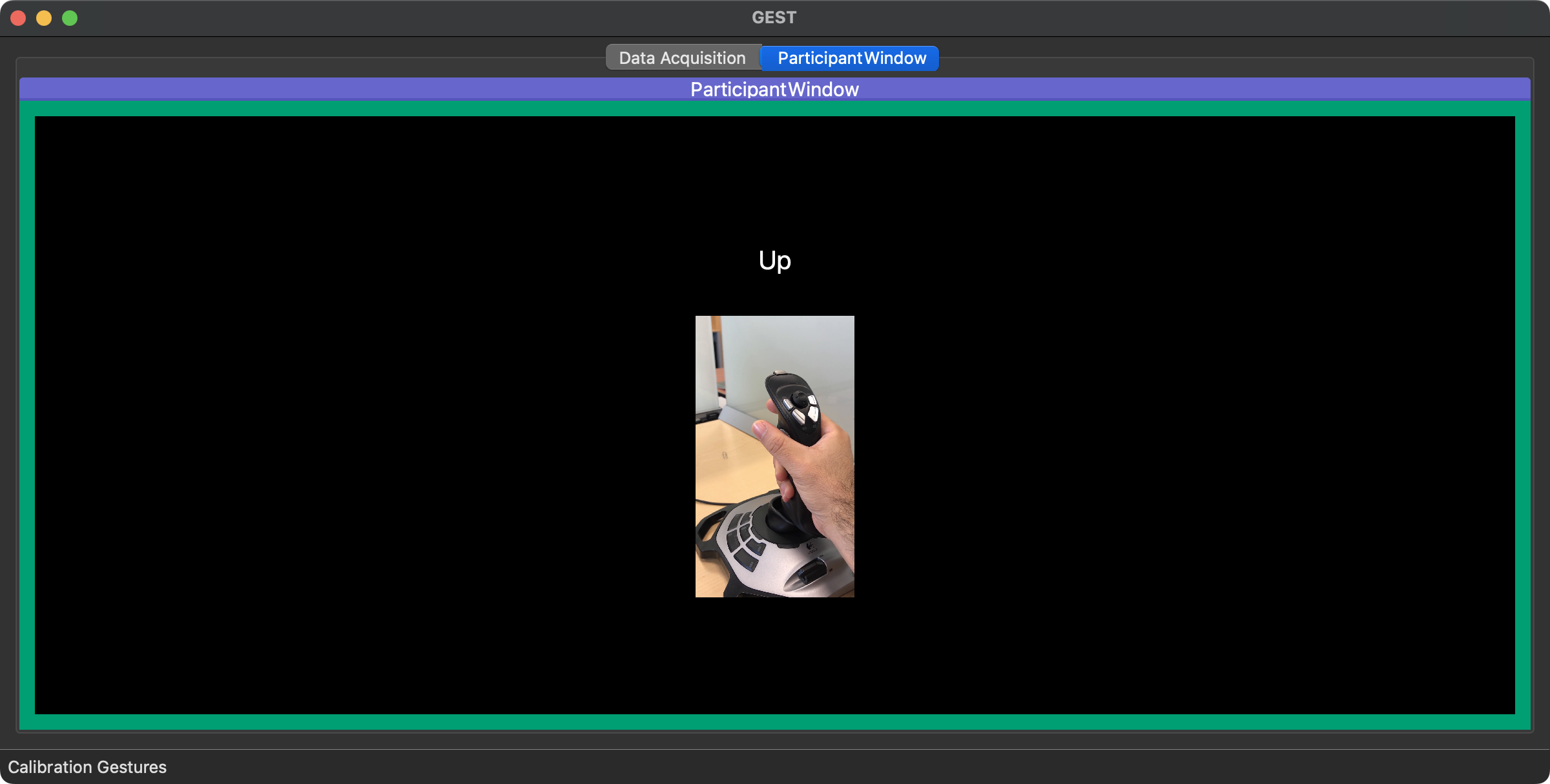}
    \vspace{1pt}
    \includegraphics[width=0.485\textwidth]{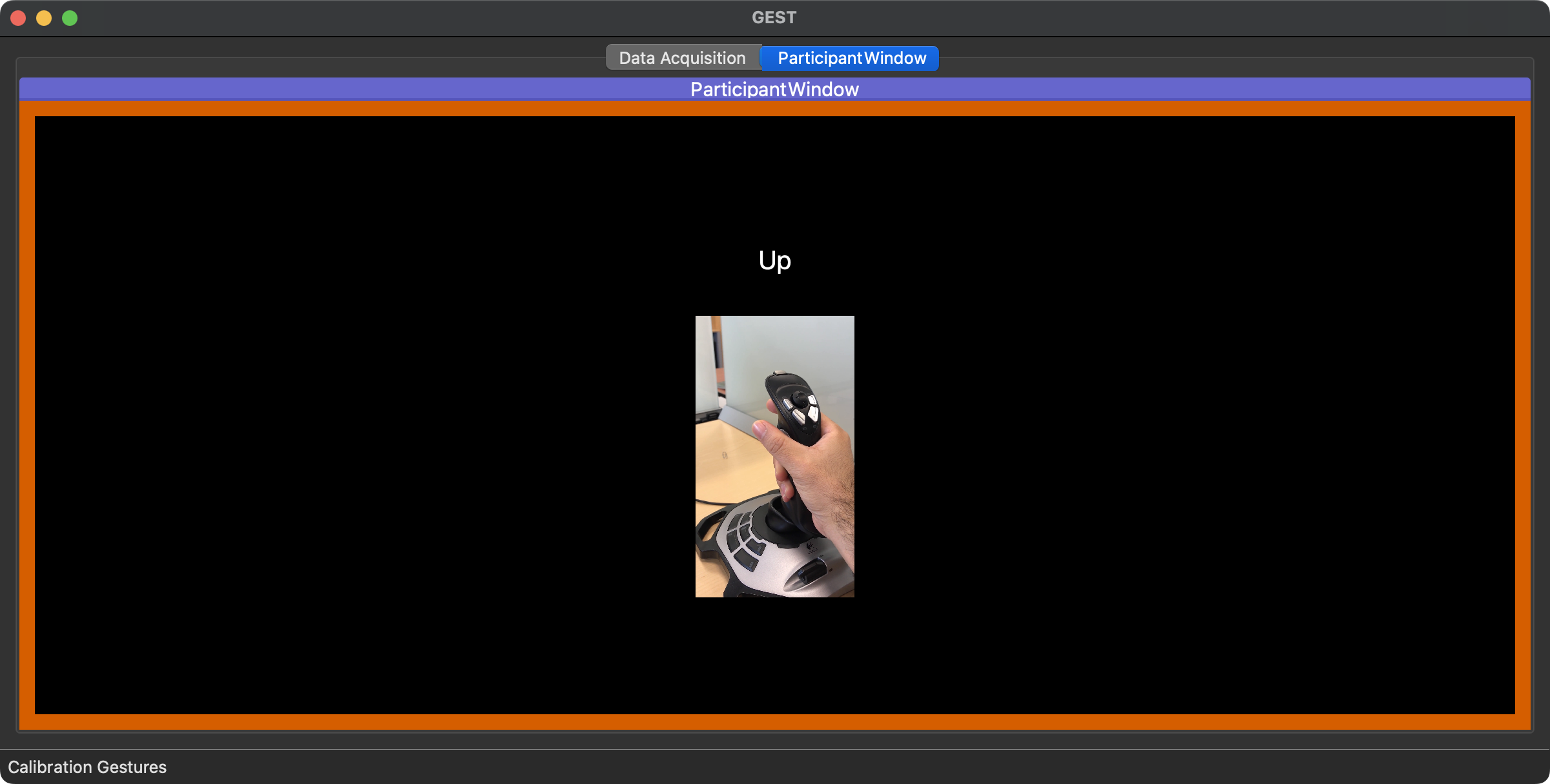}
    \caption{UI during a calibration trial. The colored screen border instructs the subject to prepare (Yellow), then hold the gesture (Green), and then rest (Red).}
    \label{fig:calibration-ui}
\end{figure}

Subjects performed an initial \textbf{Calibration} block, consisting of $8$ repetitions of each of the $7$ single gestures \lstinline{Up},
\lstinline{Down}, 
\lstinline{Left},
\lstinline{Right},
\lstinline{Pinch},
\lstinline{Thumb},
and \lstinline{Rest}.
On each calibration trial subjects were prompted to prepare for 3 seconds by a yellow screen border, then prompted to perform the gesture continuously for 2 seconds by a green border, and finally prompted to rest for 3 seconds by a red screen border. See Figure~\ref{fig:calibration-ui} for examples.

Subjects then performed alternating blocks of \textbf{Hold-Pulse} gestures (HP, one gesture component was held while the other was pulsed) and \textbf{Simultaneous-Pulse} gestures (SP, both gesture components started and stopped together). 

\begin{figure}[!ht]
    \centering
    \includegraphics[width=0.485\textwidth]{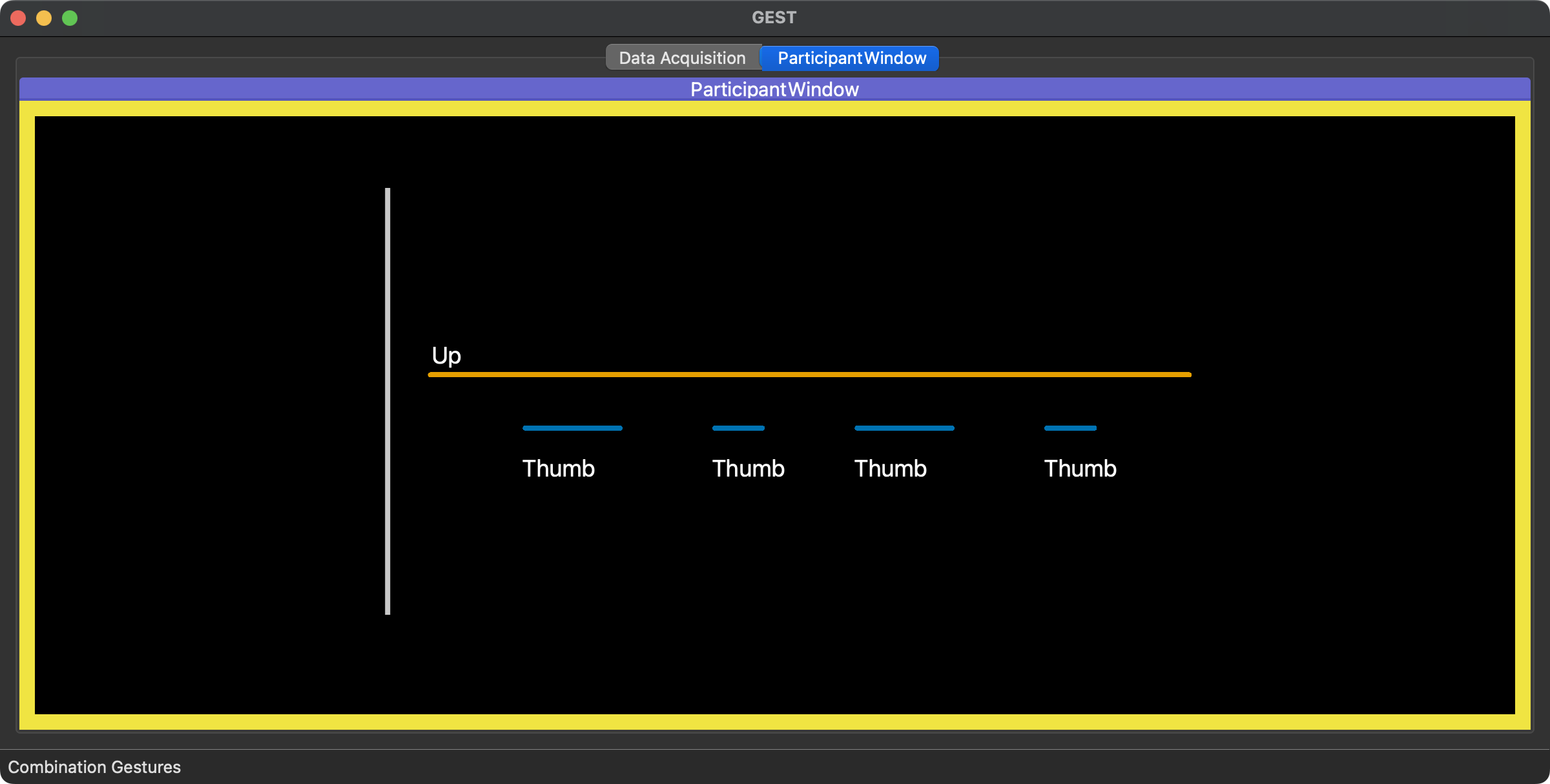}
    \caption{Subject UI during a Hold-Pulse trial. The gray vertical cursor scrolls from left to right; when it intersects a horizontal line segment, the subject performs that gesture. The screen border color also indicates when the subject should be active. }
    \label{fig:combos-ui}
\end{figure}

HP blocks contained $28$ trials, consisting of: $4 \cdot 2$ trials with held direction and pulsed modifier, $2 \cdot 4$ trials with held modifier and pulsed direction, $4$ trials with only a held direction, $2$ with only a held modifier, $4$ with only a pulsed direction, and $2$ with only a pulsed modifier. Thus a single HP block explored the full repertoire of combinations. Figure~\ref{fig:combos-ui} shows an example of the user interface (UI) shown to subjects during HP blocks. A single segment on top represents a held gesture, and $4$ shorter segments below represent another pulsed gesture.
In some trials, only the ``held'' or only the ``pulsed'' segments were present.
When a gesture trial begins, the subject's vertical cursor moves from left to right across the screen.
The gesture trial can be broken into the following stages:
\begin{itemize}
    \item The cursor starts $2$ seconds away from the instruction segments; during this time, the window border is yellow to indicate that the subject should plan their actions.
    \item For the next 8 seconds, the cursor overlaps the active area. The window border is green to indicate that the subject should perform actions as instructed; when the cursor bar intersects a colored line segment, the subject must perform that gesture. 
    \item For the final $2$ seconds, the cursor travels past the active area, and the window border are red to indicate the subject should rest. 
\end{itemize}

\begin{figure}[!ht]
    \centering
    \includegraphics[width=0.485\textwidth]{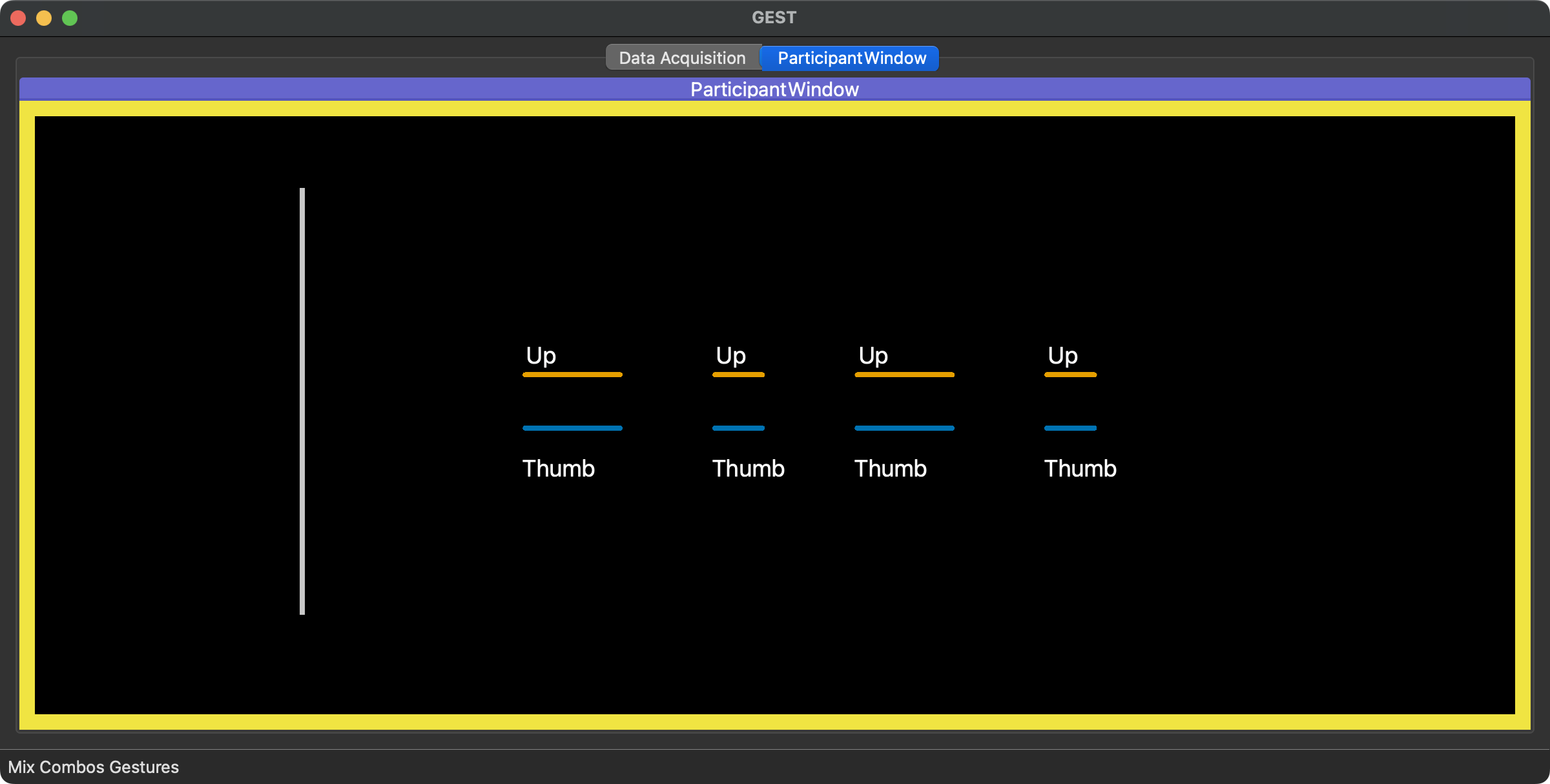}
    \caption{UI during a Simultaneous-Pulse (SP) trial. Similar to Hold-Pulse trials (Figure~\ref{fig:combos-ui}), but both gesture components onset together.}
    \label{fig:mixcombos-ui}
\end{figure}

Each Simultaneous-Pulse (SP) block contained $8$ trials; each SP trial consisted of a pair of one direction and one modifier gestures. Figure~\ref{fig:mixcombos-ui} shows an example of the UI for SP blocks. 
Unlike the HP blocks, the horizontal instruction lines were arranged so that both gesture components are started and stopped together.

Only examples of single gestures during Calibration were used for model training; any instances where the subject accidentally performed a combination gesture during this time were excluded. SP blocks contributed only examples of combination gestures, while HP blocks contributed examples of both single and combination gestures (since there were moments when only one gesture was active).

Table~\ref{tab:experiment_structure} lists the order of experimental blocks performed and the number of trials in each block. 
\begin{table}
\centering
\small
\begin{tabular}{|c|c|}
\hline
\textbf{Block Type} & \textbf{\# Gestures} \\ \hline
\hline
Calibration & 56 \\ \hline 
\hline
HP & 28 \\ \hline
SP & 8  \\ \hline
HP & 28 \\ \hline
SP & 8  \\ \hline
HP & 28 \\ \hline
SP & 8  \\ \hline 
\end{tabular}
\caption{
Experiment Structure.
Subjects performed an initial calibration, demonstrating each single gesture type multiple times. Subjects then performed multiple blocks of combination gestures. HP - Hold-Pulse one gesture was held while the other was pulsed intermittently (see Figure~\ref{fig:combos-ui}). SP - Simultaneous-Pulse; both the direction and modifier component start and end together (see Figure~\ref{fig:mixcombos-ui}).}
\label{tab:experiment_structure}
\end{table}

\subsection{Data Acquisition Framework}
\begin{figure}[!ht]
    \centering    
    \includegraphics[width=0.39\textwidth]{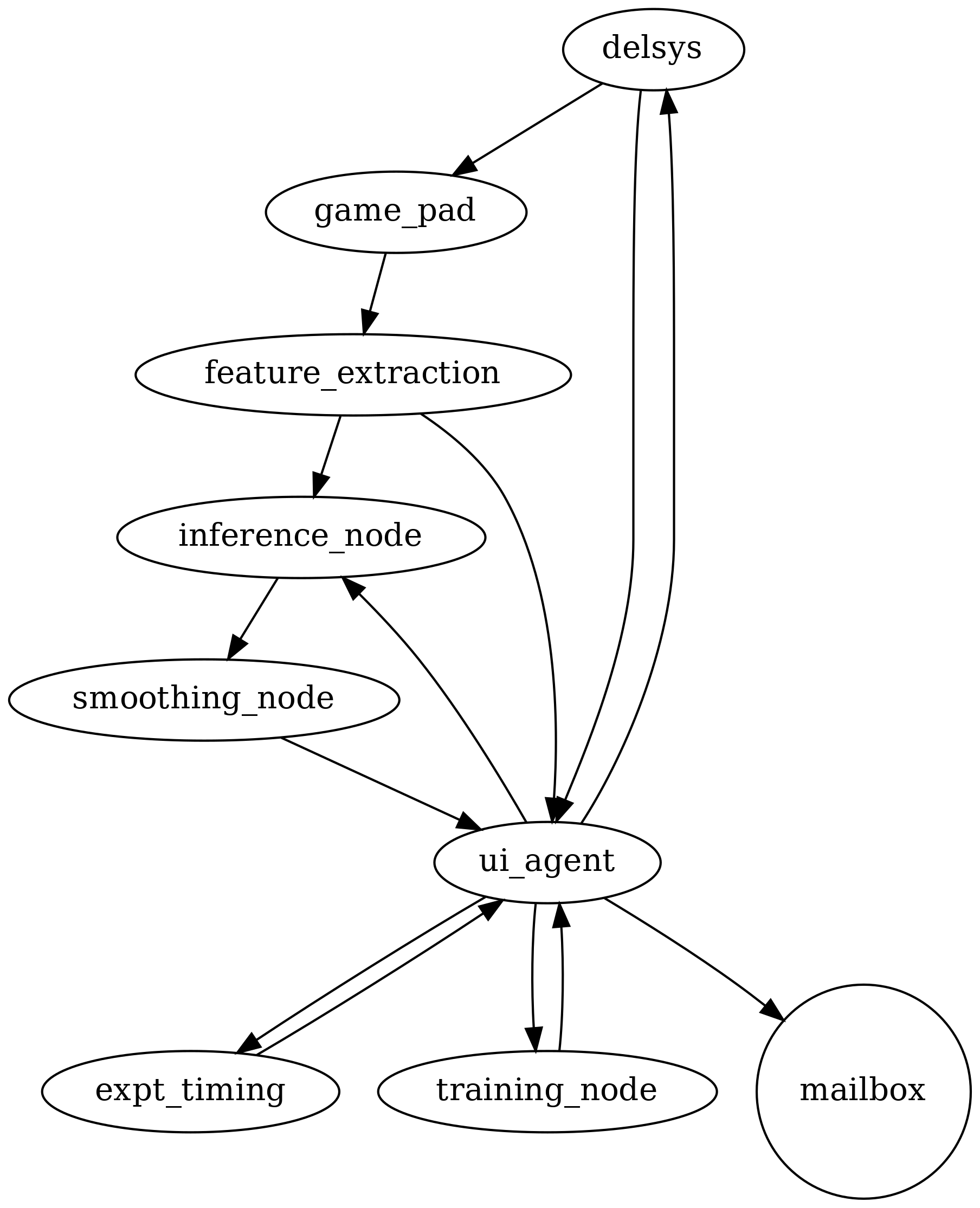}
    \caption{Flow diagram for real-time data acquisition pipeline. Each circle represents a LabGraph node with a specific responsibility; arrows indicate the flow of information.}
    \label{fig:graph}
\end{figure}

We constructed our experimental framework using the LabGraph~\cite{labgraph2021} Python package. The software was developed by creating LabGraph nodes, which handle different tasks such as collecting raw data, extracting features, training a model, performing classification and smoothing, managing the user interface, governing the experimental timing, and logging as seen in Figure~\ref{fig:graph}. 

\subsection{Structured Labels}
We defined gestures consisting of a direction component and a modifier component - this allowed us to treat these two components independently. Furthermore, we attempted to calibrate models using labeled data from only the $7$ classes of single-gesture data and extrapolate to the remaining $8$ classes of combination gestures.

In order to describe both single and combination gestures, we defined a structured label of the form \lstinline{(D,M)} consisting of a direction component and a modifier component. 
The direction component took one of $5$ values (\lstinline{Up}, \lstinline{Down}, \lstinline{Left}, \lstinline{Right}, and \lstinline{NoDir}), while the modifier component took one of $3$  values (\lstinline{Pinch}, \lstinline{Thumb}, and \lstinline{NoMod}). All $15$ classes of subject gesture behavior were described using these structured labels as follows:
\begin{itemize}
    \item Gestures with only a direction component were labeled as \lstinline{(D, NoMod)}, where \lstinline{D} was one of \lstinline{Up}, \lstinline{Down}, \lstinline{Left}, or \lstinline{Right},
    \item Gestures with only a modifier component were labeled as \lstinline{(NoDir, M)}, where \lstinline{M} was one of \lstinline{Pinch}, or \lstinline{Thumb},
    \item Gestures with both components active simply took the form \lstinline{(D, M)} where \lstinline{D} was one of the four directions and \lstinline{M} was one of the two modifiers,
    \item Resting data was labeled as \lstinline{(NoDir, NoMod)}.
\end{itemize}

\subsection{Creating Synthetic Combination Gestures}
\label{sec:synth-data}
In order to calibrate models using only single gesture examples, we derived a simple method of constructing synthetic combination gestures.
Given a feature vector from a direction gesture $z_d$ (such as \lstinline{(Up, NoMod)}) and a feature vector from a modifier gesture $z_m$ (such as \lstinline{(NoDir, Pinch)}), we constructed a synthetic feature vector $\tilde{z} = \frac{z_d + z_m}{2}$ representing an estimate of the features from the unseen combination gesture (such as \lstinline{(Up, Pinch)}).
This averaging strategy is one of the simplest possible approaches, and is similar to some approaches for data augmentation in the literature, such as Sample Pairing~\cite{inoue2018data}.

\subsection{Examining Feature Similarity}
\label{sec:examining-features-methods}
In order to explore the feature-space structure of our real and synthetic data, we constructed heatmaps showing the average similarity of items from various pairs of classes. 
In this analysis, we considered all $15$ classes of real data, and the $8$ classes of synthetic doubles gestures, giving a heatmap of shape $23$ by $23$.
We included all real single and combination gesture items, as well as a uniform subset of $0.5\%$ of all possible combination gesture items in this analysis.

We used a radial basis function (RBF) kernel to compute similarities.
Given a pair of feature vectors $z_1$ from class $C_1$ and $z_2$ from class $C_2$, the RBF kernel similarity was computed as
\begin{align}
    K_{\textsc{RBF}}(z_1, z_2) & = \exp(\shortminus\gamma \| z_1 - z_2 \|^2),
    \label{eq:rbf}
\end{align}
where the length scale $\gamma$ determines how quickly similarity should decrease as items move farther apart. 
We set the length scale $\gamma$ using the so-called ``median heuristic'': within each subject, we set $\gamma = 1 / H$, where $H$ is the median of squared Euclidean distances between any pair of feature vectors. This heuristic tends to select a length scale that gives a good contrast between similar and dissimilar items~\cite{median-heuristic}.

To compute the $(i, j)$ entry of the heatmap representing the similarity $S_{\textsc{RBF}}(C_1, C_2)$ between two classes $C_1$ and $C_2$, we averaged the RBF kernel similarity over all pairs of items in those classes,
\begin{align}
    S_{\textsc{RBF}}(C_1, C_2) & = \frac{1}{|C_1|} \frac{1}{|C_2|} \sum_{z_1 \in C_1} \sum_{z_2 \in C_2} K_{\textsc{RBF}}(z_1, z_2).
    \label{eq:avg-rbf-similarity}
\end{align}
After computing the similarity heatmap for each subject, we then averaged across subjects to obtain a single final heatmap.
Note that our similarity metric is closely related to commonly studied graph cut measures such as the normalized cut~\cite{shi2000normalized}.
The results of this feature similarity analysis are discussed in Section~\ref{sec:examining-features-results}.

\subsection{Experiment 1: Model Selection}
\label{sec:experiment_1_methods}
We performed a computational experiment to determine which choice of the model architecture and classifier algorithm would perform best in conjunction with our strategy for synthesizing combination gestures. The results of this experiment are presented in Section~\ref{sec:experiment_1_results}.

We designed all sEMG signal models to take one input vector (the features $z$ of a combination gesture) and produce two output vectors (the estimated posterior probability distribution over directions $p(y_{\textsc{dir}} | z)$, and the estimated posterior over modifiers $p(y_{\textsc{mod}} | z)$). 

In each run of the experiment, we trained and tested a model using data from a single subject.
The subject's data was divided into three parts; the Calibration data (which contained only single gestures), a test set (the final HP and SP blocks, which contained both single and combination gestures), and a special set (the other HP and SP blocks).
For all models, the test set was kept constant, and we considered three cases for how to handle the training set; 
we trained a \textbf{lower bound} model using only the single gestures from Calibration, an \textbf{upper bound} model on the single gestures from Calibration plus additional real single and combination gestures from the special set, and an \textbf{augmented} model using the real single gestures from Calibration plus synthetic combination gestures as described in Section~\ref{sec:synth-data}.
We repeated this procedure $3$ times for each subject using a different random seed.

We considered two basic model architectures, each designed around the principle that the two gesture components (direction and modifier) may be treated independently. The first approach, which we refer to as \textbf{Parallel}, is shown in Figure~\ref{fig:model-arch-ParallelA}. Here, the incoming data item is simultaneously given as input to two independent classifiers; a $5$-way classifier that was trained using only the direction component of the labels, and $3$-way classifier that was trained using only the modifier component of the labels. 
Consider the $3$-way classifier; since it used only the modifier component of the labels, it observed label values \lstinline{Pinch}, \lstinline{Thumb} and \lstinline{NoMod}. When it observed a \lstinline{NoMod} value, this could have come from any of the four direction-only gestures, or the Rest gesture. Analogously, when the $5$-way classifier observed a training item whose direction component was \lstinline{NoDir}, this item could have come from one of the two modifier-only gestures, or the Rest gesture.

In Figure~\ref{fig:model-arch-ParallelB}, we show the \textbf{Hierarchical} model architecture considered. Here, we also have two independent paths. The direction path begins with a binary classifier, predicting $p(y_{\textsc{dir}} \ne \emptyset | x)$ to answer the question: ``Is there a direction gesture present?" Then, it applies a $4$-way classifier to get the conditional probabilities of each active gesture: $p(y_{\textsc{dir}} | y_{\textsc{dir}} \ne \emptyset)$. The same two stages are taken for modifiers, with a binary classifier predicting $p(y_{\textsc{mod}} \ne \emptyset | x)$, followed by with a $2$-way classifier to get the conditional probabilities of each active gesture: $p(y_{\textsc{mod}} | y_{\textsc{mod}} \ne \emptyset)$.

We considered three choices of classification algorithm; Logistic Regression (LogR) was used as a standard method that learns linear decision boundaries, Multi-layer Perceptron (MLP) was used as a standard method that learns non-linear decision boundaries, and Random Forest (RF) was used as a representative ``rule-based'' classifier.
In each single experiment, all model components used the same classifier algorithm.
All classifiers were implemented in Python using Scikit-Learn~\cite{scikit-learn}.

Models were evaluated using balanced accuracy, which is the average accuracy in each class. We further separated performance into the balanced accuracy on single gestures (``Singles Acc'', i.e. the $7$ classes for which real data was available from Calibration), the balanced accuracy on double gestures (``Doubles Acc'', i.e. the $8$ unseen gesture classes in which both label components are active), and the overall balanced accuracy (``Overall Acc'').
\begin{figure}[!ht]
\centering
\begin{tikzpicture}[
square/.style={rectangle, draw=black, very thick},
shadedsq/.style={rectangle, draw=blue, fill=blue!5, thin},
arrow/.style={-latex, draw=black, thick} 
]
    \begin{pgfonlayer}{fg}    
        \node[square]   (Input) at (1, 0) {Input};
        \node[square]   (ClassifyDirection) at (-1, -2) {ClassifyDirection};
        \node[shadedsq] (Up) at (-3, -4) {U};
        \node[shadedsq] (Down) at (-2, -4) {D};
        \node[shadedsq] (Left) at (-1, -4) {L};
        \node[shadedsq] (Right) at (0, -4) {R};
        \node[shadedsq] (NoDir) at (1, -4) {$\emptyset$};
        \node[square]   (ClassifyModifier) at (3, -2) {ClassifyModifier};
        \node[shadedsq] (Thumb) at (2, -4) {T};
        \node[shadedsq] (Pinch) at (3, -4) {P};
        \node[shadedsq] (NoMod) at (4, -4) {$\emptyset$};
    \end{pgfonlayer}
    \draw [style=arrow, out=270, in=90] (Input) to (ClassifyDirection);
    \draw [style=arrow, out=270, in=90] (ClassifyDirection) to (Up);
    \draw [style=arrow, out=270, in=90] (ClassifyDirection) to (Down);
    \draw [style=arrow, out=270, in=90] (ClassifyDirection) to (Right);
    \draw [style=arrow, out=270, in=90] (ClassifyDirection) to (NoDir);
    \draw [style=arrow, out=270, in=90] (ClassifyDirection) to (Left);
    \draw [style=arrow, out=270, in=90] (Input) to (ClassifyModifier);
    \draw [style=arrow, out=270, in=90] (ClassifyModifier) to (Thumb);
    \draw [style=arrow, out=270, in=90] (ClassifyModifier) to (Pinch);
    \draw [style=arrow, out=270, in=90] (ClassifyModifier) to (NoMod);
    \begin{pgfonlayer}{mid}
        \node[fill=white, draw, rounded corners, inner sep=0.1cm, thick, drop shadow, 
        fit=(Up) (Down) (Left) (Right) (NoDir)
    ] {};
    \end{pgfonlayer}
    \begin{pgfonlayer}{mid}
        \node[fill=white, draw, rounded corners, inner sep=0.1cm, thick, drop shadow, 
        fit=(Thumb) (Pinch) (NoMod)
    ] {};
    \end{pgfonlayer}
    \begin{pgfonlayer}{bg}
        \node[fill=white, draw, rounded corners, inner sep=0.25cm, thick, drop shadow, 
        fit=(Input) (ClassifyDirection) (Up) (Down) (Left) (Right) (NoDir) (ClassifyModifier) (Thumb) (Pinch) (NoMod)       
    ] {};
    \end{pgfonlayer}
\end{tikzpicture}
\caption{
\textbf{Parallel} Model Architecture. Input data is processed by two independent components. The first component predicts probabilities among $5$ classes representing directions: \textbf{U}p, \textbf{D}own, \textbf{L}eft, \textbf{R}ight and ``NoDirection" ($\emptyset$). The second component predicts probabilities among $3$ modifier classes representing: \textbf{T}humb, \textbf{P}inch, and ``NoModifier" (\textbf{$\emptyset$}).
}
\label{fig:model-arch-ParallelA}
\vspace{5mm}
\begin{tikzpicture}[
square/.style={rectangle, draw=black, very thick, in front of path},
shadedsq/.style={rectangle, draw=blue, fill=blue!5, thin, in front of path},
arrow/.style={-latex, draw=black, thick} 
]
    \node[square]   (Input) at (1, 2) {Input};
    \node[square]   (DetectDirection) at (-0.8, 0) {DetectDirection};
    \node[square]   (ClassifyDirection) at (-2, -2) {ClassifyDirection};
    \node[shadedsq] (Up) at (-3, -4) {U};
    \node[shadedsq] (Down) at (-2, -4) {D};
    \node[shadedsq] (Left) at (-1, -4) {L};
    \node[shadedsq] (Right) at (0, -4) {R};
    \node[shadedsq] (NoDir) at (1, -4) {$\emptyset$};
    \node[square]   (DetectModifier) at (3, 0) {DetectModifier};
    \node[square]   (ClassifyModifier) at (2, -2) {ClassifyModifier};
    \node[shadedsq] (Thumb) at (2, -4) {T};
    \node[shadedsq] (Pinch) at (3, -4) {P};
    \node[shadedsq] (NoMod) at (4, -4) {$\emptyset$}; 
    \draw [style=arrow, out=270, in=90] (Input) to (DetectDirection);
    \draw [style=arrow, out=270, in=90] (DetectDirection) to (NoDir);
    \draw [style=arrow, out=180, in=120] (Input) to (ClassifyDirection);
    \draw [style=arrow, out=270, in=90] (DetectDirection) to (ClassifyDirection);
    \draw [style=arrow, out=270, in=90] (ClassifyDirection) to (Up);
    \draw [style=arrow, out=270, in=90] (ClassifyDirection) to (Down);
    \draw [style=arrow, out=270, in=90] (ClassifyDirection) to (Left);
    \draw [style=arrow, out=270, in=90] (ClassifyDirection) to (Right);
    \draw [style=arrow, out=0, in=90] (Input) to (DetectModifier);
    \draw [style=arrow, out=270, in=90] (DetectModifier) to (NoMod);
    \draw [style=arrow, out=270, in=120] (Input) to (ClassifyModifier);
    \draw [style=arrow, out=270, in=90] (DetectModifier) to (ClassifyModifier);
    \draw [style=arrow, out=270, in=90] (ClassifyModifier) to (Thumb);
    \draw [style=arrow, out=270, in=90] (ClassifyModifier) to (Pinch);
    \begin{pgfonlayer}{mid}
        \node[fill=white, draw, rounded corners, inner sep=0.1cm, thick, drop shadow, 
        fit=(Up) (Down) (Left) (Right) (NoDir)
    ] {};
    \end{pgfonlayer}
    \begin{pgfonlayer}{mid}
        \node[fill=white, draw, rounded corners, inner sep=0.1cm, thick, drop shadow, 
        fit=(Thumb) (Pinch) (NoMod)
    ] {};
    \end{pgfonlayer}
    \begin{pgfonlayer}{bg}
        \node[fill=white, draw, rounded corners, inner sep=0.25cm, thick, drop shadow, 
        fit=(Input) (DetectDirection) (ClassifyDirection) (Up) (Down) (Left) (Right) (NoDir) (DetectModifier) (ClassifyModifier) (Thumb) (Pinch) (NoMod)       
    ] {};
    \end{pgfonlayer}
    
\end{tikzpicture}
\caption{\textbf{Hierarchical} Model Architecture. Input data is processed in two independent components. 
In the first component, 
a binary classifier determines the probability that a direction is absent ($\emptyset$); then, a $4$-way classifier is used to determine the conditional probability over directions: \textbf{U}p, \textbf{D}own, \textbf{L}eft, \textbf{R}ight, given that a direction is present. 
Analogous processing occurs in the modifier component.
}
\label{fig:model-arch-ParallelB}
\end{figure}

\subsection{Experiment 2: Selecting Synthetic Combination Gestures}
\label{sec:experiment_2_methods}
Based on the results of Experiment 1, we selected the ``Parallel'' model architecture with MLP classifier algorithm for further analysis. 
Our next experiment explored different ways to use subsets of all possible synthetic combination gestures. The results of this experiment are presented in Section~\ref{sec:experiment_2_results}.

As expected, we observed during the first experiment that adding synthetic combination gestures was necessary in order for models to successfully classify combination gestures at test time.
In Experiment 1, we created synthetic double gestures by naively combining all possible valid pairs (consisting of features from one direction-only gesture and features from one modifier-only gesture); this resulted in a training dataset where the majority of data was synthetic.
We observed that the augmented model, and even some of the upper-bound models, experienced a trade-off between the performance on single gestures and the performance on double gestures.
We hypothesize that this trade-off might be partly due to the over-abundance of synthetic data; therefore, in our next experiment, we explored ways of reducing the amount of synthetic data to try to improve this trade-off.

Consider the case of constructing synthetic combination feature vectors $\tilde{z}$ from two classes of single gestures $C_d$ and $C_m$ (e.g. \lstinline{(Up, NoMod)} and \lstinline{(NoDir, Pinch)}).
In Experiment 1, when constructing all possible pairs, we create all $|C_d| |C_m|$ possible items.
Now in Experiment 2, we considered two main approaches: subsetting the two single gesture classes \textit{first}, and then creating all possible pairs (methods prefixed by \textit{subsetInput\_}), or creating all possible pairs and then selecting a subset \textit{afterwards} (methods prefixed by \textit{subset\_}). 

For each of these approaches, we considered three ways of subsetting items:  
taking a representative sample by sampling items uniformly at random (suffixed by \_\textit{uniform}),
taking a tightly clustered sample by selecting items closest to the mean (suffixed by \_\textit{near}\_\textit{mean}),
and taking a diverse sample by selecting items whose quantile of distance to the mean was evenly spaced (e.g. to take $4$ points, we would take the points whose quantile of distance to the mean were $q=0.0, 0.33, 0.66, 1.0$) (suffixed by \_\textit{spaced}\_\textit{quantiles}).

We included the same lower bound and upper bound models in this experiment.
The ``baseline'' for this model was the performance of an augmented model using all possible synthetic items.
As with Experiment 1, models were evaluated for their balanced accuracy on single gesture classes, double gesture classes, or overall across all classes, and experimental runs were repeated for $3$ random seeds for each subject.

\subsection{Experiment 3: Single Gesture Data Augmentation}
\label{sec:experiment_3_methods}
In Experiment 2, we found that creating all possible synthetic combinations and then taking a uniform random subset of $10\%$ nearly preserved performance while greatly reducing the amount of synthetic data used during training. 
However, we still observed a trade-off between the performance on single gestures and the performance on combination gestures. 
Thus, in Experiment 3, we explored methods for adding augmented single gesture data in the hope of improving this trade-off between accuracy on single gestures and double gestures.
The results of this experiment are presented in Section~\ref{sec:experiment_3_results}.

Based on Experiment 2, we selected a single setting for further exploration: a uniform subset of $10\%$ after creating all synthetic pairs.
We then considered three basic methods for creating augmented single gesture features from real single gesture features. First, we considered adding Gaussian noise to real items (\textit{add-gaussian-*}), where noise $\epsilon$ was sampled $\epsilon \sim \mathcal{N}(0, \sigma^2 I)$ with $\sigma$ values of $0.3$, $0.4$, or $0.5$. Next, we considered fitting a Gaussian mixture model (GMM) to the data and sampling points from this estimated distribution (\textit{fit-gmm-*}); we considered GMMs with $1$, $5$, or $10$ components. Lastly, we considered fitting the distribution of features using a kernel density estimate (KDE), and sampling points from this estimated distribution (\textit{fit-kde}). For this KDE approach, we used a Gaussian kernel with a fixed lengthscale of $0.01$. Fitting a GMM and fitting a KDE were both done using Scikit-Learn~\cite{scikit-learn}.

We again considered the lower bound and upper bound models as described in the previous two experiments. The baseline for each part of Experiment 3 was the performance of a model trained with synthetic combination gestures and no augmented single gestures.

\section{Results}
\subsection{Examining Feature Similarity}
\label{sec:examining-features-results}
\begin{figure*}[!ht]
    \centering
    \includegraphics[width=0.8\textwidth]{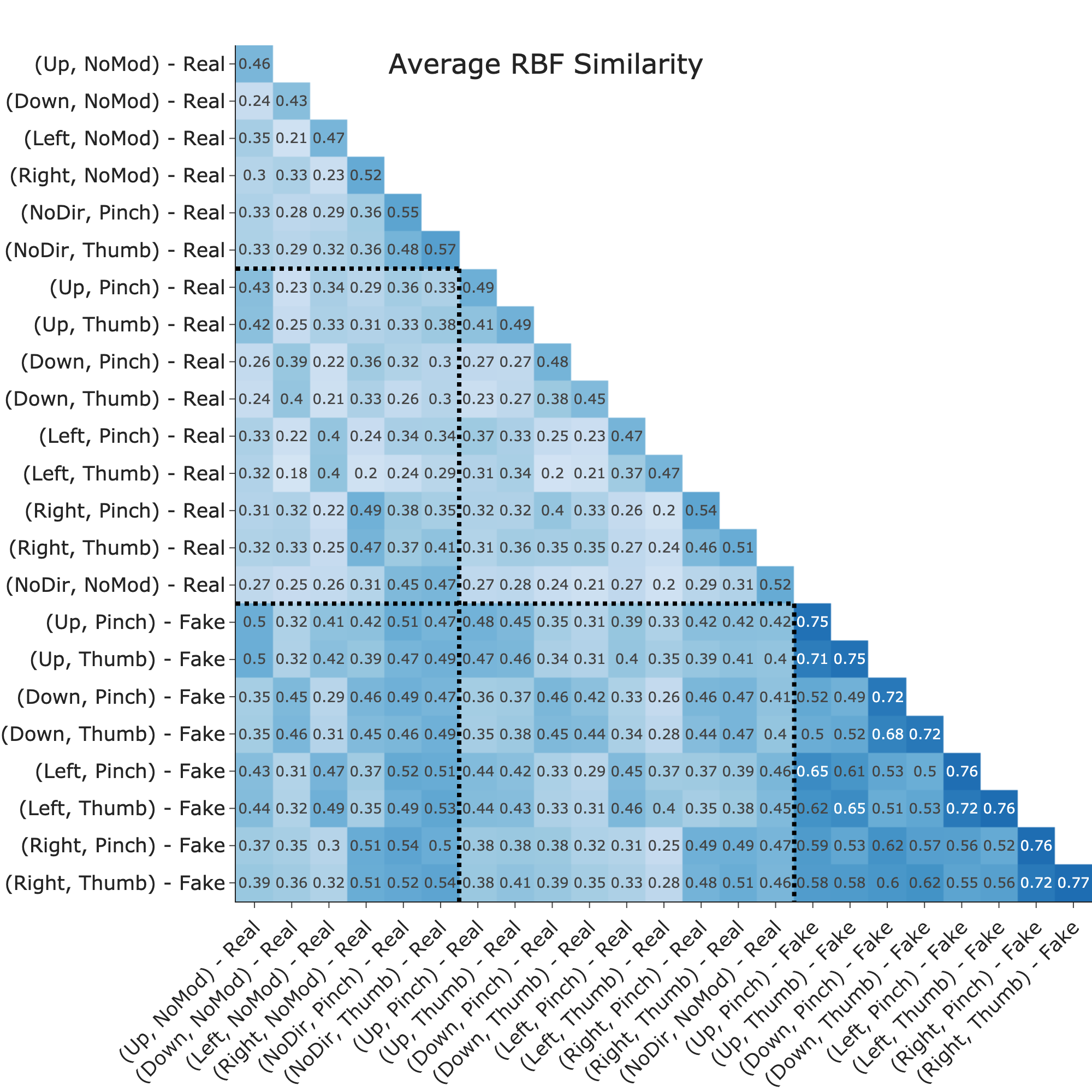}
    \caption{Feature Similarity Heatmap. Using pre-computed features from real single gestures, real combination gestures, and synthetic combination gestures, we computed average pairwise similarity $S_{\textsc{RBF}}(C_1, C_2)$ between pairs of classes according to Eq.~\ref{eq:avg-rbf-similarity}. After computing values within each subject, we averaged across subjects to obtain a single heatmap.
    }
    \label{fig:heatmap}
\end{figure*}

In Figure~\ref{fig:heatmap}, we show the average pairwise RBF kernel similarity between the features of different gesture classes.
These similarities were computed using Eq.~\ref{eq:avg-rbf-similarity}, as described in Section~\ref{sec:examining-features-methods}.
The dotted guidelines on the heatmap indicate distinct regions of interest (comparison between real singles, real combinations, and synthetic combinations).

Note that, for an individual pair of feature vectors, the RBF kernel described in Eq.~\ref{eq:rbf} takes values ranging from $0$ when items have very dissimilar features, to $1$ when items have identical features. 
The average pairwise similarity between classes described in Eq.~\ref{eq:avg-rbf-similarity} therefore also takes values in the same range.
Values on the diagonal of the heatmap represent the similarity of items within a class and can be viewed as a measure of the spread of a certain class in feature space.

Before interpreting the results we observed, it is useful to consider what we expect to see in this similarity heatmap in an ideal case, where subjects performed gestures very precisely (i.e. every repetition of a certain gesture is performed the same way), and we chose an ideal feature space that accurately represents the structure of the raw data. 
In this case, we may hope to observe that the similarity heatmap has high values on the diagonal, indicating that items belonging to the same class have very similar features.
For different classes of single gestures (such as \lstinline{Up} and \lstinline{Pinch}), if we believe the underlying movements are distinct, we may hope to observe that the feature similarity is low.
    When comparing a single gesture (such as \lstinline{Up}) to a combination gesture it belongs to (such as \lstinline{Up, Pinch}), it is difficult to make \textit{a priori} predictions about the similarity, but we may hope that these class pairs are closer than other class pairs that share no gesture components.
    Finally, if we chose a suitable method for creating synthetic combination gestures, then we may hope to observe that their features are highly similar to the matching real classes.
This would appear as a diagonal line in the heatmap, in the region comparing real and synthetic items.

In Figure~\ref{fig:heatmap}, we observed many of the expected trends.
The main diagonal is stronger, indicating that each class is relatively tightly clustered.
In the middle-left section, comparing real double gestures and real single gestures, we observe that \lstinline{(D, M)} is highly similar to \lstinline{(D, NoMod)} for each direction gesture \lstinline{D}, but not very similar to the modifier component \lstinline{M}; this indicates that the feature vectors are more strongly influenced by their direction component.
We observe a similar trend in the bottom-left section, comparing synthetic doubles to real singles.
In the bottom-middle section, comparing synthetic doubles to real doubles, we see some similarity between the expected class pairings, such as synthetic \lstinline{(Up, M)} and real \lstinline{(Up, M)}, but we also see a similarity between unexpected class pairings, such as synthetic \lstinline{(Up, M)} and real \lstinline{(Left, M)}.
In the bottom-right section, comparing synthetic doubles to synthetic doubles, we again see that the direction component drives the structure of the feature vectors; as a result, \lstinline{(D, M1)} and \lstinline{(D, M2)} are similar for any modifiers \lstinline{M1} and \lstinline{M2}.

\subsection{Experiment 1: Model Selection}
\label{sec:experiment_1_results}
\begin{figure}[!ht]
    \centering
    \includegraphics[width=0.495\textwidth]{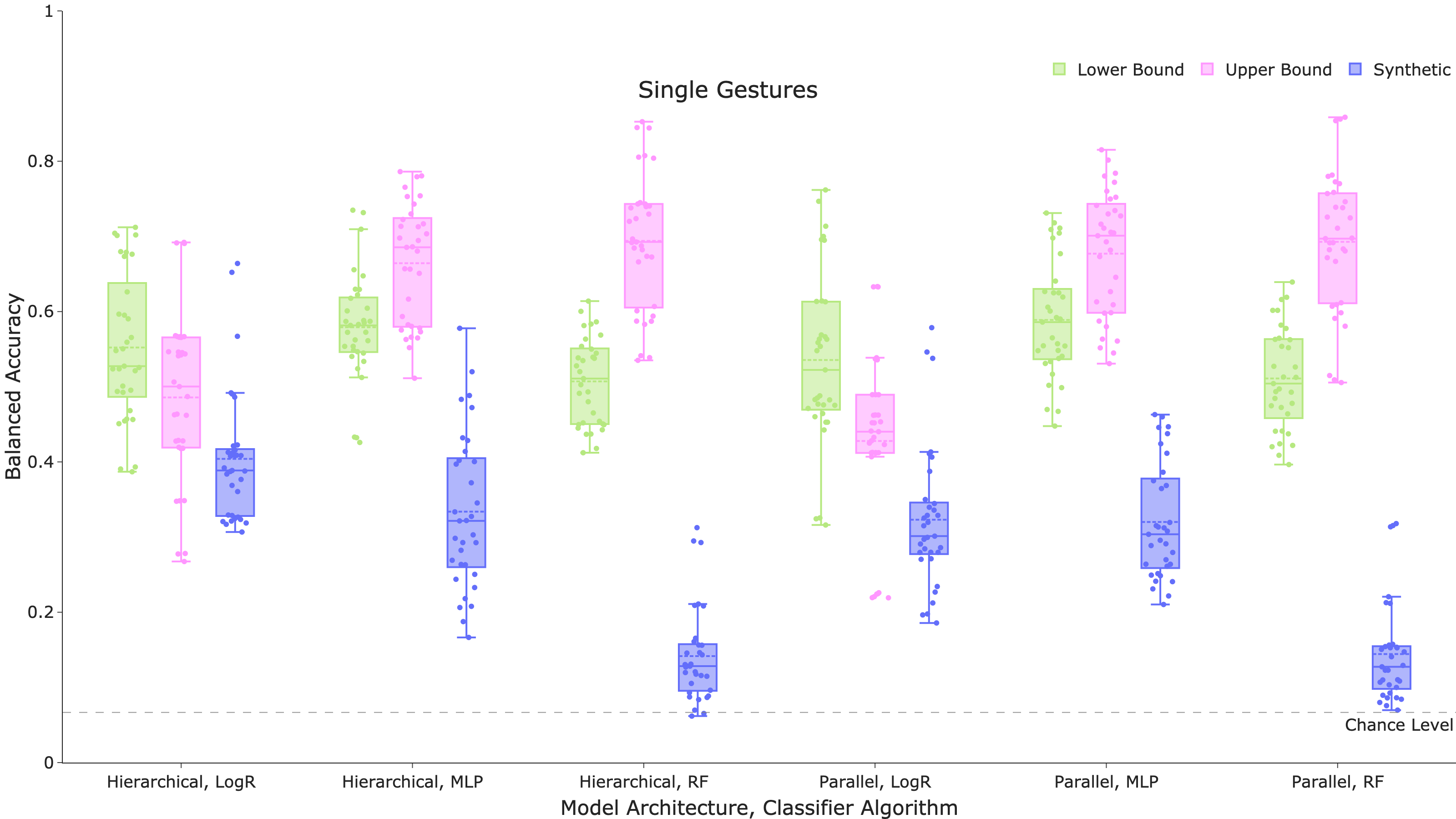}
    \includegraphics[width=0.495\textwidth]{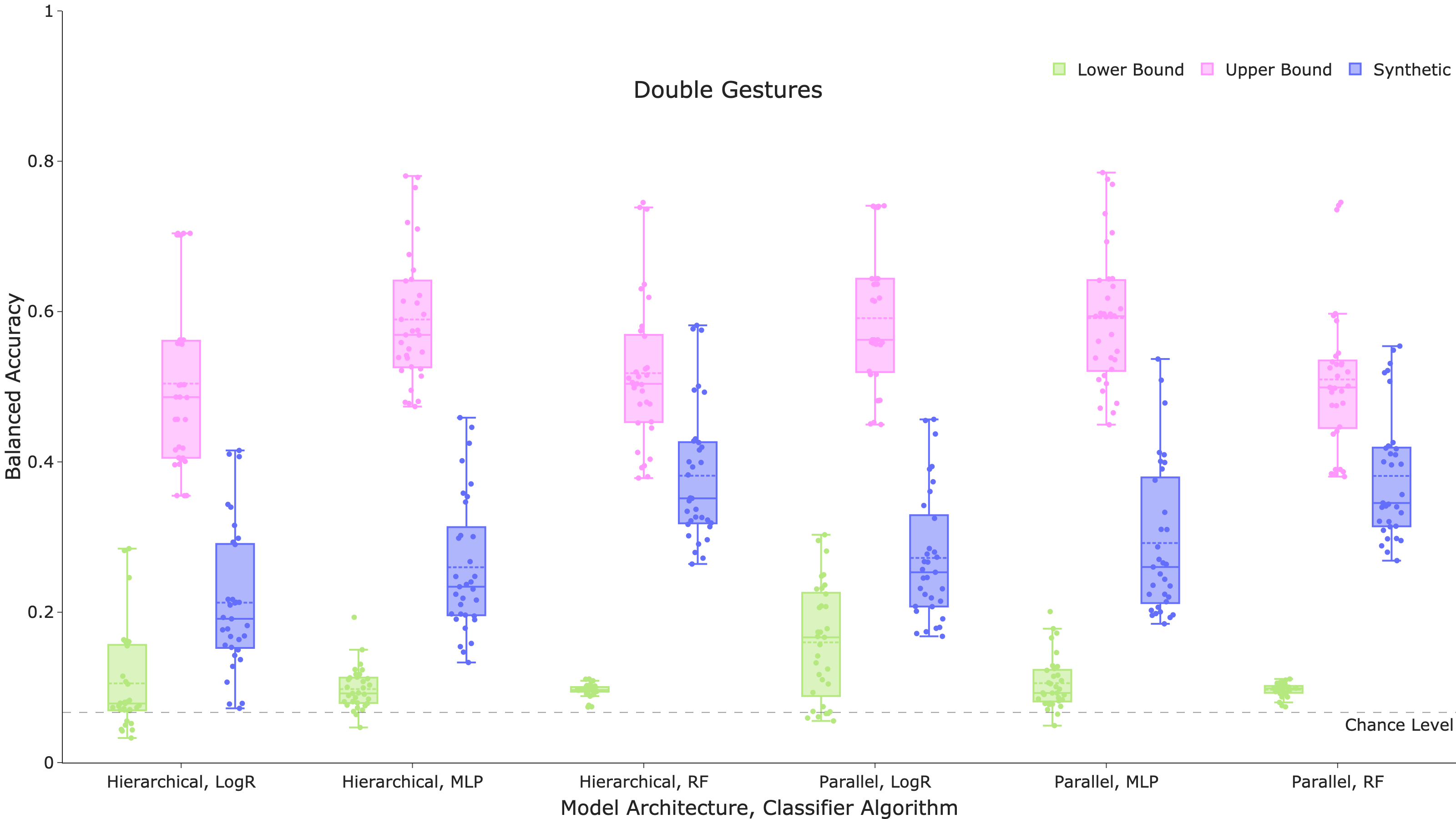}
    \includegraphics[width=0.495\textwidth]{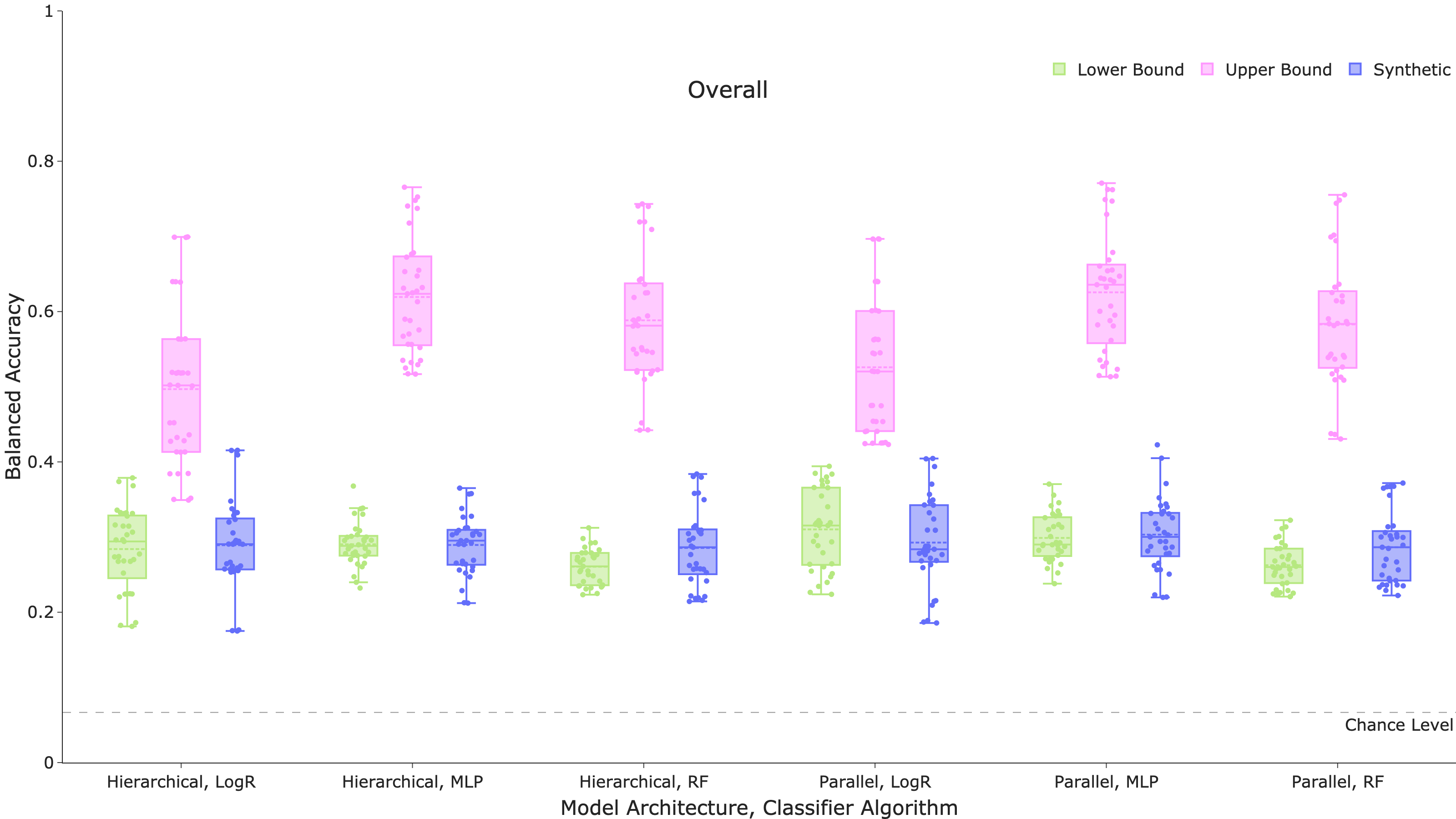}
    \caption{Experiment 1: balanced accuracy using various classifier algorithms (\textit{LogR} - logistic regression, \textit{MLP} - multi-layer perceptron, and \textit{RF} - Random Forest) and model architectures (\textit{Parallel} and \textit{Hierarchical}). 
    Top - performance on single gesture classes; Middle - performance on double gesture classes; Bottom - performance on all gesture classes.
    \textit{Lower Bound} model is trained on Calibration data only; \textit{Synthetic} model is trained on Calibration data with synthetic double gestures (using $100\%$ of the generated doubles); \textit{Upper Bound} model is trained on Calibration, HP1, HP2, SP1, and SP2 data. All models are tested on HP3 and SP3 data. Each dot represents one subject and one random seed.}
    \label{fig:experiment1}
\end{figure}
In Figure~\ref{fig:experiment1}, we show the results of Experiment 1, in which we varied the choice of model architecture and classification algorithm. See Section~\ref{sec:experiment_1_methods} for a detailed explanation of this experiment. The top panel of Figure~\ref{fig:experiment1} shows balanced accuracy on single gestures only, while the middle panel shows balanced accuracy on double gestures only, and the bottom panel shows overall balanced accuracy for all gesture classes.

We observed the best trade-off between accuracy on single gestures and double gestures using the MLP classifier with Parallel architecture. Although the Random Forest classifier performed best for double gestures classification, it performed significantly worse than other methods in single gestures classification. Since the Parallel architecture performed better for double gesture classification with a small trade-off in single gesture classification, we selected the MLP classifier with Parallel architecture for further exploration in the next two experiments.

\subsection{Experiment 2: Selecting Synthetic Combination Gestures}
\label{sec:experiment_2_results}
Using the best classifier and architecture from \ref{sec:experiment_1_results}, MLP with Parallel architecture, we explored methods to subset synthetic double gestures, as well as the number of synthetic double gestures to use in this experiment. See Section~\ref{sec:experiment_2_methods} for further details.

\begin{figure}[!ht]
    \centering
    \includegraphics[width=0.495\textwidth]{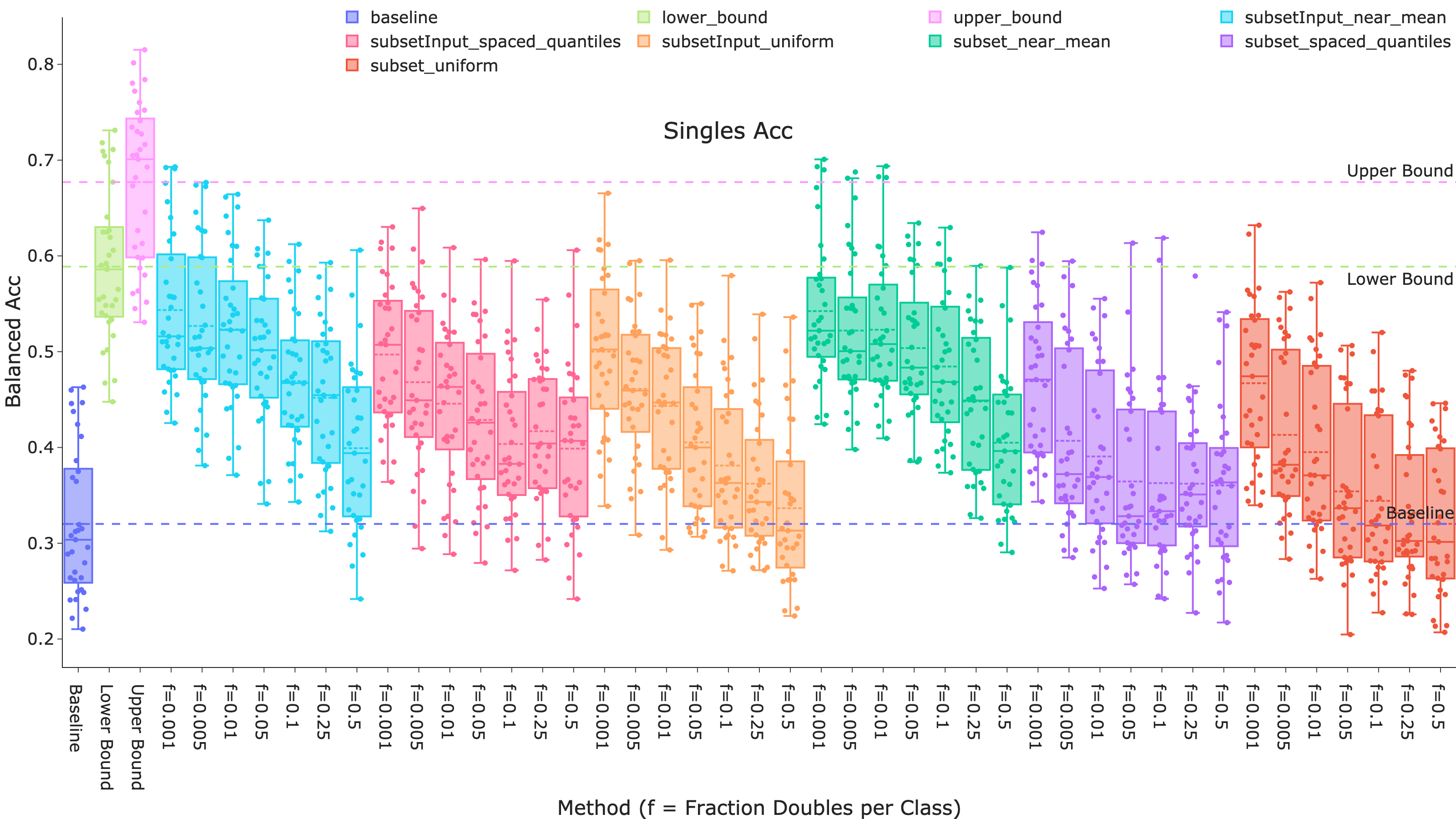}
    \includegraphics[width=0.495\textwidth]{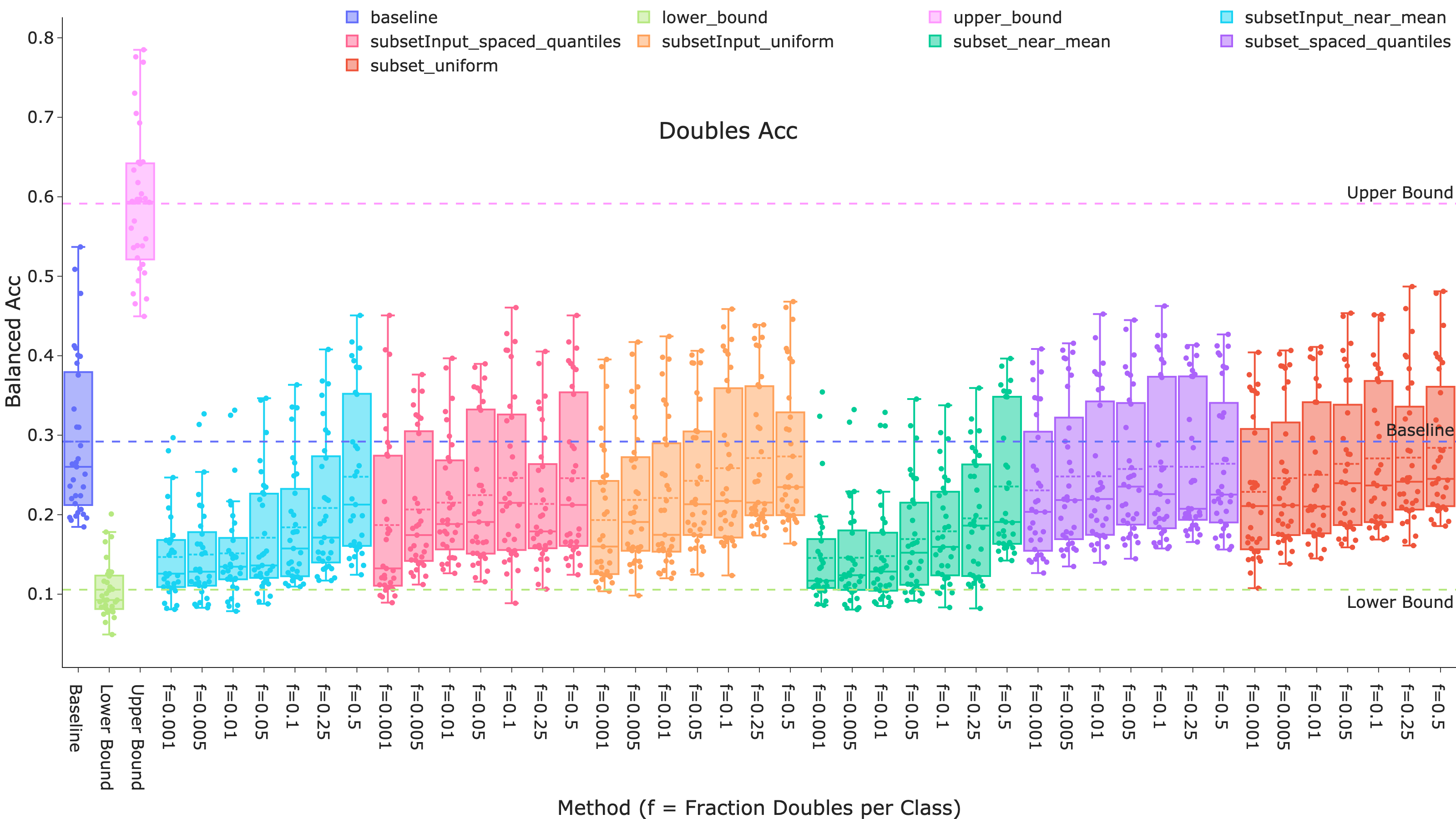}
    \includegraphics[width=0.495\textwidth]{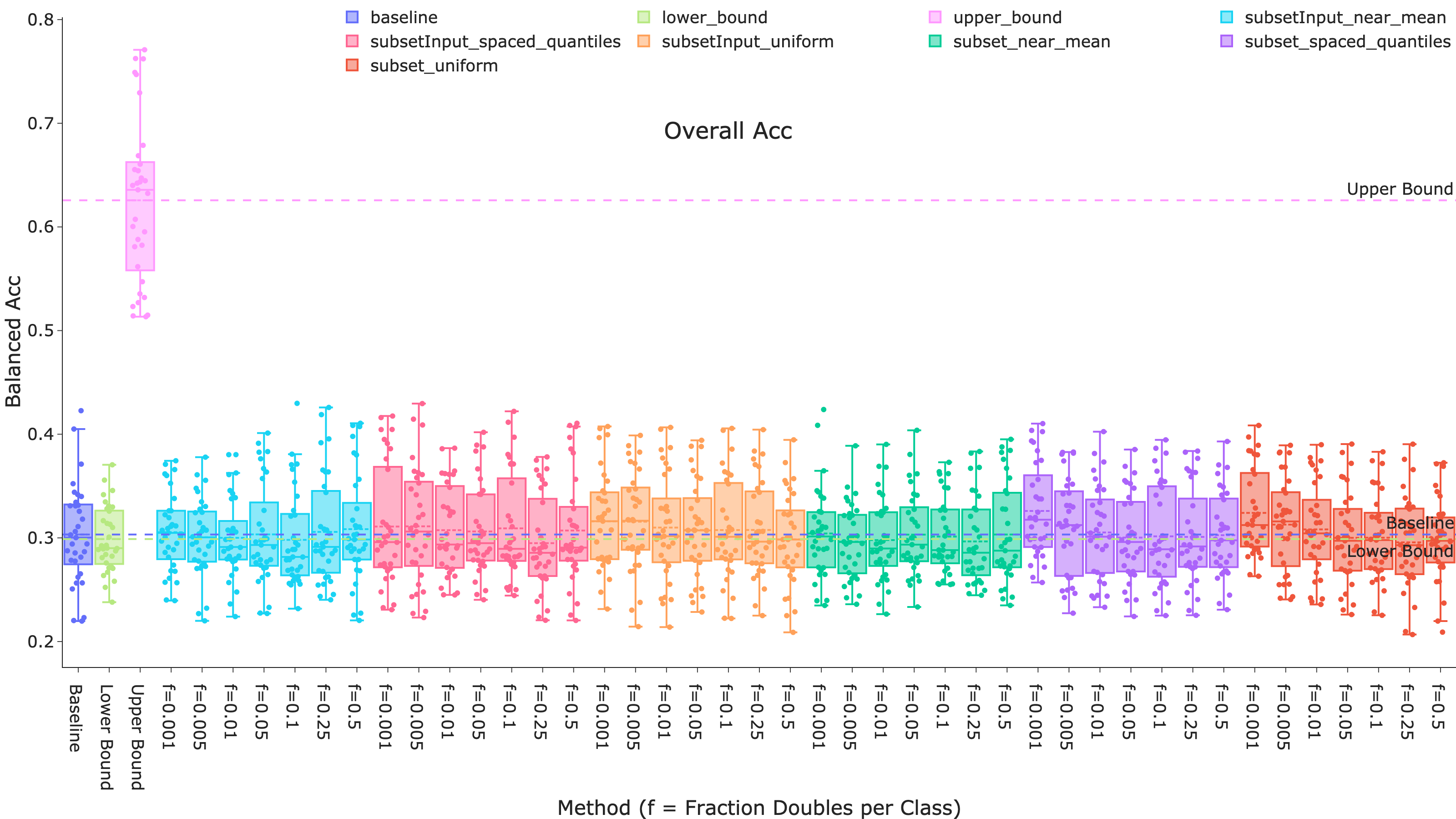}
    \caption{
    Experiment 2: balanced accuracy using \textit{MLP} classifier and \textit{Parallel} architecture with various strategies to select subsets of synthetic double gestures.
    Top - performance on single gesture classes;
    Middle - performance on double gesture classes;
    Bottom - performance on all gesture classes.
    \textit{Lower Bound} model trained on Calibration data only;
    \textit{Upper Bound} model trained on Calibration, HP1, HP2, SP1, and SP2 data;
    \textit{Baseline} model trained on Calibration and all synthetic data;
    Other models trained on Calibration and subsets of synthetic data ($f \in \{ 0.001, 0.005, 0.01, 0.05, 0.1, 0.25, 0.5 \}$ represents the amount of synthetic data used).
    All models are tested on HP3 and SP3 data.
    Each dot represents one subject and one random seed.}
    \label{fig:experiment2}
\end{figure}
In Figure~\ref{fig:experiment2}, we show the results of this experiment. As before, the top panel shows balanced accuracy on single gestures, while the middle panel shows balanced accuracy on double gestures, and the bottom panel shows balanced accuracy overall. 
In general, we found that sampling diverse items (using \textit{*\_subset\_spaced\_quantiles} methods) or sampling items near the mean (using \textit{*\_subset\_near\_mean} methods) had an overly-strong effect and resulted in a trade-off; the resulting models experienced a large decrease in performance on double gestures and a large increase in performance on single gestures.
We also found that the model appeared to learn useful information from the whole set of synthetic double gestures, not just the ones near the mean. We found that taking a uniform (and therefore representative) subset gave a reasonable approximation of the same performance, but helped increase training speed and helped reduce the excess of synthetic data in our training set.

\subsection{Experiment 3: Single Gesture Data Augmentation}
\label{sec:experiment_3_results}
\begin{figure}[!ht]
    \centering
    \includegraphics[width=0.495\textwidth]{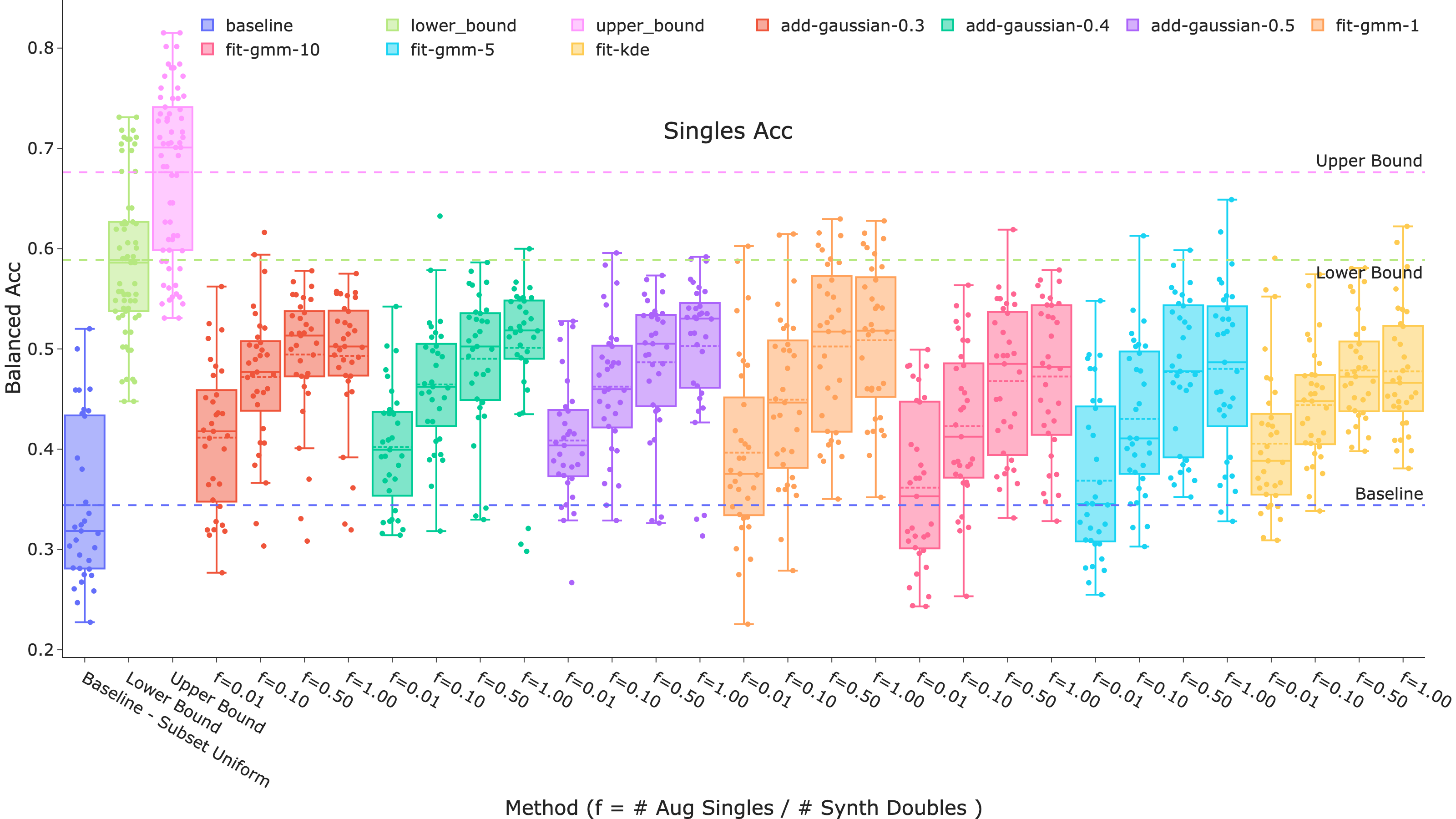}
    \includegraphics[width=0.495\textwidth]{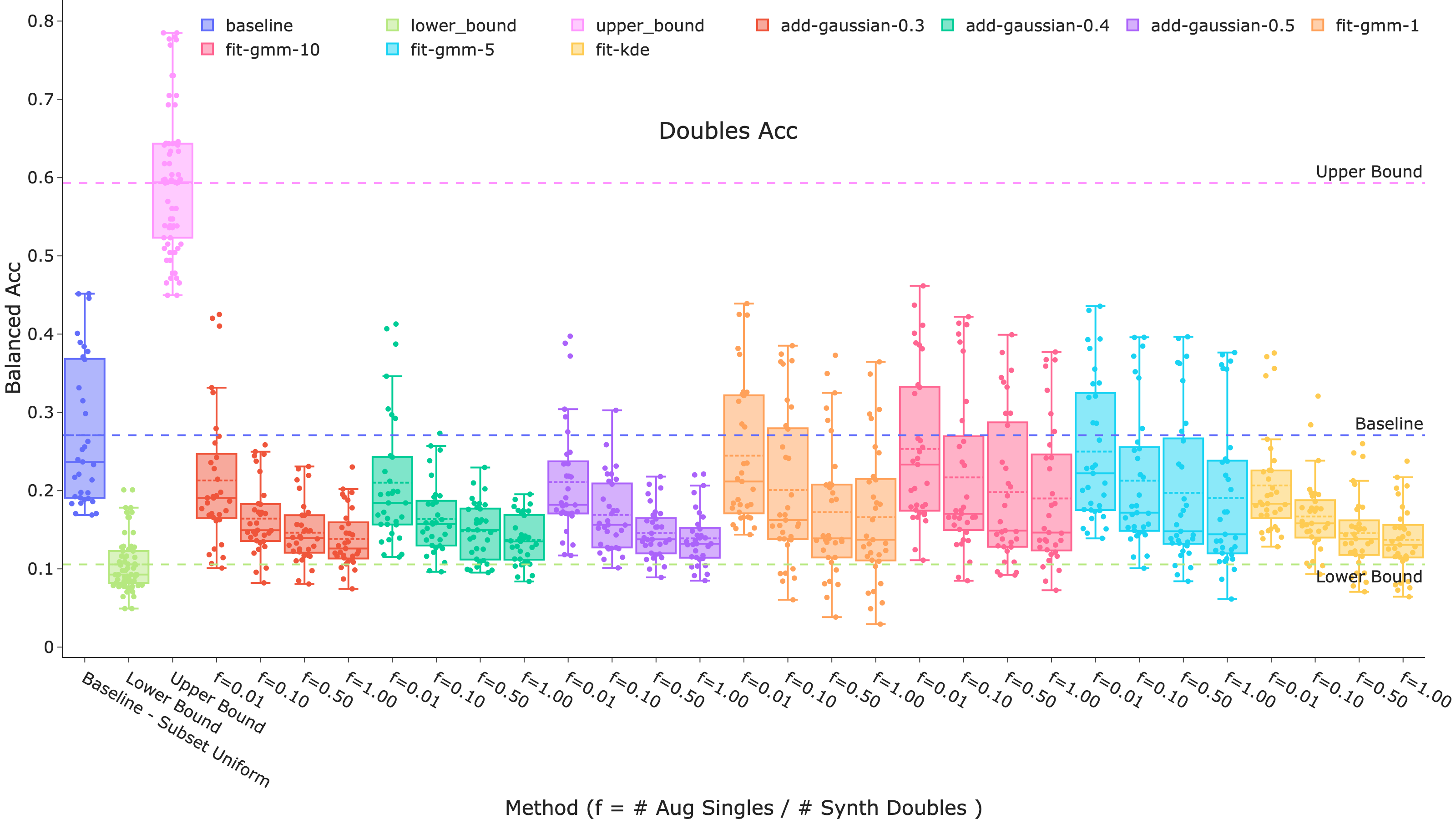}
    \includegraphics[width=0.495\textwidth]{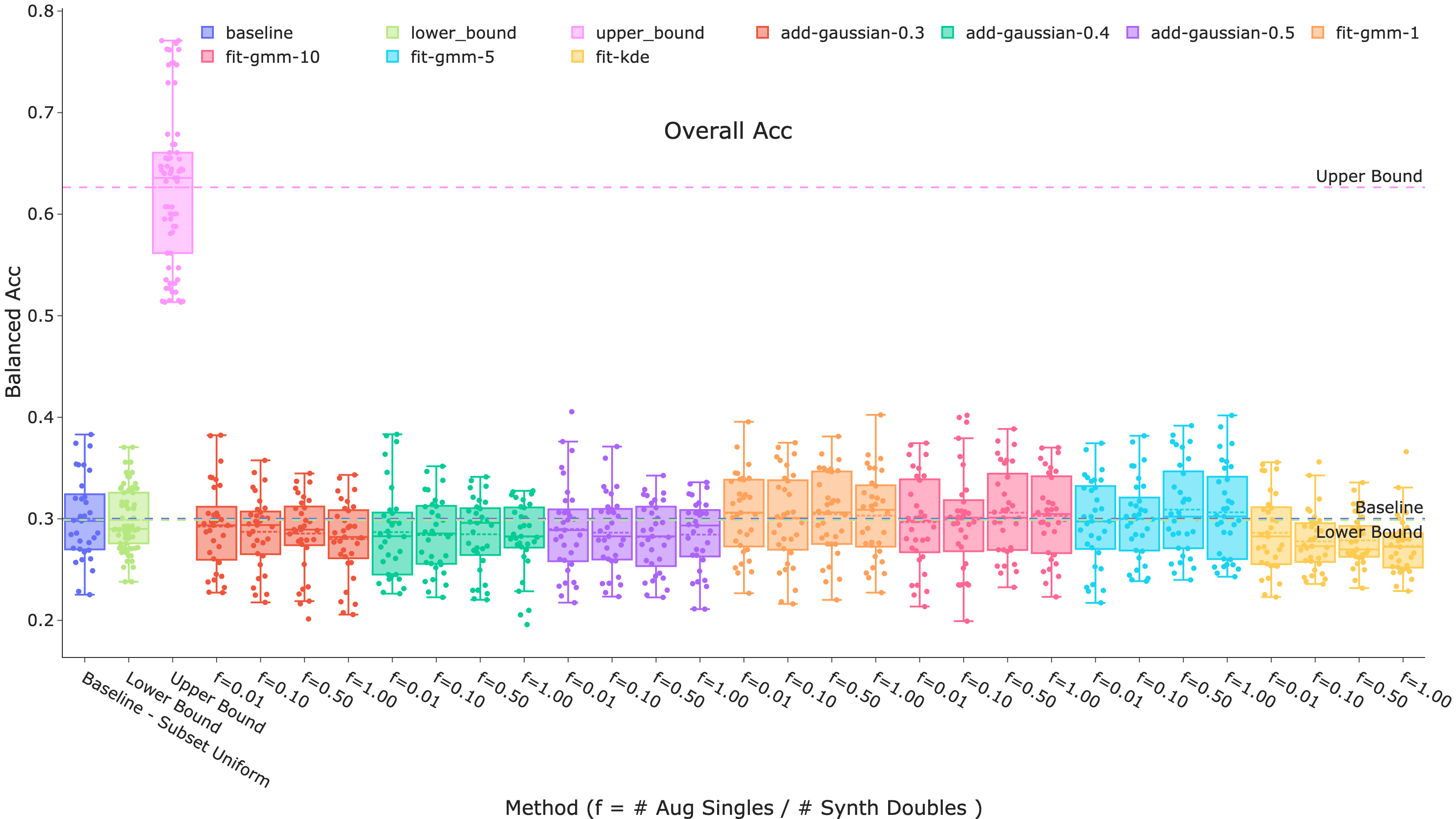}
    \caption{
    Experiment 3: balanced accuracy using augmented single gestures.
    Top - performance on single gesture classes;
    Middle - performance on double gesture classes;
    Bottom - performance on all gesture classes.
    \textit{Lower Bound} model trained on Calibration data only;
    \textit{Upper Bound} model trained on Calibration, HP1, HP2, SP1, and SP2 data;
    \textit{Baseline} model trained on Calibration and \textit{uniform} $10\%$ of synthetic data after mixing;
    Other models trained on Calibration, $10\%$ of synthetic data, and different amounts of augmented single gesture data ($f \in \{0.001, 0.005, 0.01, 0.05, 0.1, 0.25, 0.5 \} $ 
    represents the amount of augmented singles to create relative to the amount of synthetic doubles).
    All models are tested on HP3 and SP3 data.
    Each dot represents one subject and one random seed.
    }
    \label{fig:experiment3-subset}
\end{figure}
In Section~\ref{sec:experiment_2_results}, we explored the effect of using subsets of synthetic double gestures. Although we found that it was possible to roughly maintain model performance while reducing the number of synthetic items used, we still observed an undesirable trade-off between model performance on single gestures and doubles gestures.
Roughly speaking, adding more (synthetic) double gestures to the training set causes the model to achieve higher accuracy on double gestures at the cost of achieving lower accuracy on single gestures.
We hypothesize that this trade-off may be due to the imbalance between the amount of single and double gestures in the training set, and we attempt to resolve this trade-off by adding additional single gestures to the training set using data augmentation. 
As described previously, we considered several methods for creating augmented single gestures.
Figure~\ref{fig:experiment3-subset} shows the performance when the training set includes a subset of synthetic double gestures plus a set of augmented single gestures.
The subset of synthetic doubles for Figure~\ref{fig:experiment3-subset} was taken using the uniform random strategy, keeping $10\%$ of possible items for each class.

We observe that, even though we managed to increase the accuracy on singles gestures with data augmentation, this came at the cost of losing model performance on double gesture classification. Figure~\ref{fig:experiment3-subset} shows that GMM methods (\textit{fit-gmm-1}, \textit{fit-gmm-5} and \textit{fit-gmm-10}) with 10\% of synthetic doubles gave us the best trade-off between the gain of accuracy on single gestures and loss of accuracy on double gestures.
We also experimented with adding augmented single gestures alongside all possible synthetic double gestures; the trends in performance from this scenario were not substantially different from Figure~\ref{fig:experiment3-subset}

\section{Discussion}
We proposed a method for increasing the expressivity of gestures derived from sEMG signals using a novel architecture for multi-label gesture classification. The approach we describe addresses the inherent trade-off between the inclusion of an expressive gesture set and the need for exhaustive calibration through the use of synthetic data generated through the merging of single gesture feature vectors to create synthetic combinations. Given the novel nature of this approach, we report on the performance of different classifiers, and methods of selecting synthetic data to guide future developments in the field of gesture expressivity. 

\subsection{Model Selection for Gesture Expressivity}
Prior approaches to multi-label classification have primarily focused on the generation of composite gestures from individual components such as the construction of a fist gesture from the collective flexion of individuals digits~\cite{bjorklund2018investigating, mendes2022multi} or the classification of multiple axes of action of a single joint~\cite{young2012classification, hahne2015concurrent}. In contrast, we focus on combining gestures that could be used for joint actions leveraging the knowledge of limb biomechanics to construct sub groups of gestures that are mutually exclusive in their actions. To do so, we made use of a problem transformation approach to train individual classifiers for each subgroup of gestures. Each subgroup of gestures can be interpreted as discrete action sets that can be combined to form joint actions. We found that there was no advantage to the hierarchal model structure in determining whether a gesture from an action set was present prior to selecting an action from that set compared to when such a determination was made at the same time an action was selected. We also examined the performance of logistic regression, multi-layer perceptron, or random forest classifier algorithms within this model architecture. We observed better performance of the multi-layer perceptron compared to other models when considering the trade-off between the ability to classify single gestures and double gestures. This results likely reflects the highly non-linear boundaries between classes. Overall, we found that this problem transformation approach yielded promising results, and allowed us to increase the expressivity of our gesture set, while maintaining a relatively short model calibration time and also avoiding physiologically-implausible predictions. 

In our experiments, we included a lower-bound model trained on only real single gestures, as well as an upper-bound model trained using real single gestures and real double gestures. We found that, for a sufficiently flexible classifier algorithm such as MLP, the upper-bound model did \textit{not} experience a trade-off between single and double gesture performance. This may indicate that generating realistic and informative synthetic data is a key direction for future research.

\subsection{Use of Synthetic and Augmented Data}
The need to collect training data on all possible combinations of single gestures represents a limiting factor in the ability to expressivity of gestures through combination, as the calibration time needed quickly becomes laborious. We therefore imposed a restriction during our model development, that combinations of gestures were to be omitted from the training set. As seen in our lowe-bound models classification of combination gestures using only single gestures yields poor (near chance) results. We, therefore, tested the use of synthetic combination gestures, where each synthetic combination was created by averaging the feature vectors of a real direction-only gesture and a real modifier-only gesture.
Our basic approach for creating synthetic combination gestures was to generate all possible valid pairs. While this approach yielded significant improvements in the classification performance of double gestures, it also resulted in a trade-off between model performance on single and double gesture classes. We hypothesized that this was due to the extreme over-abundance of synthetic data when including all valid pairs. We therefore also explored methods of subsetting these synthetic combination gestures, and found that using a uniform random subset of $10\%$ of generated items preserved most of the performance gains while greatly reducing the amount of synthetic training data. However, even with the use of $10\%$ of generated combination items a large class imbalance between synthetic combinations and real single gestures remained. We, therefore, tested whether the augmentation of single items using different noise models would improve the decrement in singles accuracy due to the addition of doubles. We did not find this to be the case. The augmentation of singles data to reduce the class imbalance did not have a profound effect on increasing singles accuracy. 

\subsection{Limitations and Future Directions}
The experiments conducted in the current study included only a limited number of gestures, and relied on a joystick for obtaining ground truth labels. As the number of direction gesture classes and the number of modifier gesture classes increases, the number of possible combination gestures would increase combinatorially. Thus, successfully training a model that can extrapolate from the sum of component gestures to the product-set of all possible gesture combinations would yield a greater potential benefit, in terms of the amount of time saved during model calibration. However, it remains to be seen whether the current approach will scale to more single gesture classes. The current approach could also be scaled up by using ``chords'' of 3 or more gestures; this approach would be practically limited by the need for multiple sets of biomechanically independent gestures, the need for suitable hardware to obtain ground truth labels, and may offer diminishing utility in a real-world application due to the high skill requirements of using such multi-gesture chords.

Our experiments focused on the use of two-component model architectures. We also considered using a single model over all possible combination classes. One key drawback is that such a model does not have a well-defined decision rule for unseen classes; thus it cannot be used in the ``lower-bound'' control scenario. We experimented with such a single model using synthetic data and found that it did not out perform the proposed parallel and hierarchical model architectures.

In the current study, based on previous observations of the significant impact of label noise due to subject task non-adherence, we used a set of gestures that was well captured using a joystick. This gave us high confidence in the set of labels, but some of the gestures used may have relatively weaker signals in the distal-forearm sEMG sensor setup that was used.
In particular, our direction gestures consisted of gross wrist movements, which we expect to be well measured by distal-forearm sEMG, but our modifier gestures used fine-grained finger movements, which rely partially on hand-intrinsic muscle activity that may not be captured as well. Future work may explore alternative, more flexible hardware for ground-truth labeling, so that we may ensure that all gestures used are well captured by the sEMG sensor setup.

Another important direction for future research is to evaluate other methods for creating synthetic training data, including other simple functions and even functions with learnable parameters that may be pre-trained on a separate population of subjects.

Finally, an area of potential improvement is the use of deep neural networks. 
These models offer two key possible benefits. First, by using classifier models with greater capacity, we may be able to make better use of a large set of synthetic training data, or even train population models to extrapolate to unseen subjects.
Second, a gradient-based training scheme may lend itself well to several alternative training schemes. For example, we may consider a two-stage approach, in which the model is first pre-trained on a set of real single gestures, and then fine-tuned on a set of synthetic combination gestures. Alternatively, we may use a multi-term objective function that treats real and synthetic data differently.

Despite these limitations, our results indicate that the novel problem transformation approach tested using a parallel model architecture in combination with a non-linear classifier, and restricted synthetic data generation holds potential for increasing the expressivity of sEMG-based gestures with short calibration time.

\subsection{Applications of Proposed Method}
Current approaches to control of multi-joint robotic or prosthetic limbs rely on a ``direct control'' approach, in which a subject must learn to simultaneously manipulate each degree of freedom. 
This style of control yields expressivity, but requires significant user training and expertise.
The proposed methodology has the potential to provide the same expressivity while greatly reducing the burden of training for the user. 
This approach also offers a framework for greatly increasing the expressivity of human-computer interaction; the use of combination gestures allows for a combinatorial increase in decoded user actions, similar to the use of the ``shift,'' ``control,'' or ``caps lock,'' keys on a keyboard.
The proposed method may also facilitate more easily incorporating new gestures on-the-fly, since the model is explicitly designed to extrapolate to unseen combinations without exhaustive supervision.

\section{Ethical Statement}
All protocols were conducted in conformance with the Declaration of Helsinki and were approved by the Institutional Review Board of Northeastern University (IRB number 15-10-22) 
All subjects providing informed written consent before participating.

\section{Acknowledgements}
Funding for this research was provided by Meta Reality Labs Research. 

\section{References}
\bibliography{references.bib}

\begin{thebibliography}{10}

\bibitem{yang2021dynamic}
Zhiwen Yang, Ying~Sun Du~Jiang, Bo~Tao, Xiliang Tong, Guozhang Jiang, Manman
  Xu, Juntong Yun, Ying Liu, Baojia Chen, and Jianyi Kong.
\newblock Dynamic gesture recognition using surface emg signals based on
  multi-stream residual network.
\newblock {\em Frontiers in Bioengineering and Biotechnology}, 9, 2021.

\bibitem{xiong2021deep}
Dezhen Xiong, Daohui Zhang, Xingang Zhao, and Yiwen Zhao.
\newblock Deep learning for emg-based human-machine interaction: a review.
\newblock {\em IEEE/CAA Journal of Automatica Sinica}, 8(3):512--533, 2021.

\bibitem{qi2019intelligent}
Jinxian Qi, Guozhang Jiang, Gongfa Li, Ying Sun, and Bo~Tao.
\newblock Intelligent human-computer interaction based on surface emg gesture
  recognition.
\newblock {\em IEEE Access}, 7:61378--61387, 2019.

\bibitem{tsoumakas2006review}
Grigorios Tsoumakas, Ioannis Katakis, and Ioannis Vlahavas.
\newblock A review of multi-label classification methods.
\newblock In {\em Proceedings of the 2nd ADBIS workshop on data mining and
  knowledge discovery (ADMKD 2006)}, pages 99--109, 2006.

\bibitem{young2012classification}
Aaron~J Young, Lauren~H Smith, Elliott~J Rouse, and Levi~J Hargrove.
\newblock Classification of simultaneous movements using surface emg pattern
  recognition.
\newblock {\em IEEE Transactions on Biomedical Engineering}, 60(5):1250--1258,
  2012.

\bibitem{hahne2015concurrent}
Janne~M Hahne, Sven D{\"a}hne, Han-Jeong Hwang, Klaus-Robert M{\"u}ller, and
  Lucas~C Parra.
\newblock Concurrent adaptation of human and machine improves simultaneous and
  proportional myoelectric control.
\newblock {\em IEEE Transactions on Neural Systems and Rehabilitation
  Engineering}, 23(4):618--627, 2015.

\bibitem{olsson2021learning}
Alexander~E Olsson, Neboj{\v{s}}a Male{\v{s}}evi{\'c}, Anders Bj{\"o}rkman, and
  Christian Antfolk.
\newblock Learning regularized representations of categorically labelled
  surface emg enables simultaneous and proportional myoelectric control.
\newblock {\em Journal of NeuroEngineering and Rehabilitation}, 18:1--19, 2021.

\bibitem{olsson2019extraction}
Alexander~E Olsson, Paulina Sager, Elin Andersson, Anders Bj{\"o}rkman,
  Neboj{\v{s}}a Male{\v{s}}evi{\'c}, and Christian Antfolk.
\newblock Extraction of multi-labelled movement information from the raw
  hd-semg image with time-domain depth.
\newblock {\em Scientific reports}, 9(1):1--10, 2019.

\bibitem{bjorklund2018investigating}
Petter Bj{\"o}rklund.
\newblock Investigating the use of multi-label classification methods for the
  purpose of classifying electromyographic signals.
\newblock Master's thesis, Lund University, 2018.

\bibitem{mendes2022multi}
Jos{\'e} Jair~Alves Mendes~Junior, Carlos~Eduardo Pontim, and Daniel~Prado
  Campos.
\newblock Multi-label emg classification of isotonic hand movements: A suitable
  method for robotic prosthesis control.
\newblock In {\em Brazilian Congress on Biomedical Engineering}, pages
  1665--1671. Springer, 2022.

\bibitem{hermens1992median}
Hermanus~J Hermens, TAMv Bruggen, Christian~TM Baten, WLC Rutten, and HBK Boom.
\newblock The median frequency of the surface emg power spectrum in relation to
  motor unit firing and action potential properties.
\newblock {\em Journal of Electromyography and Kinesiology}, 2(1):15--25, 1992.

\bibitem{labgraph2021}
Jimmy Feng, Pradeep Damodara, George Gensure, Ryan Catoen, and Allen Yin.
\newblock Labgraph.
\newblock \url{https://github.com/facebookresearch/labgraph}, 2021.

\bibitem{inoue2018data}
Hiroshi Inoue.
\newblock Data augmentation by pairing samples for images classification.
\newblock {\em arXiv preprint arXiv:1801.02929}, 2018.

\bibitem{median-heuristic}
Damien Garreau, Wittawat Jitkrittum, and Motonobu Kanagawa.
\newblock Large sample analysis of the median heuristic.
\newblock {\em arXiv preprint arXiv:1707.07269}, 2017.

\bibitem{shi2000normalized}
Jianbo Shi and Jitendra Malik.
\newblock Normalized cuts and image segmentation.
\newblock {\em IEEE Transactions on pattern analysis and machine intelligence},
  22(8):888--905, 2000.

\bibitem{scikit-learn}
F.~Pedregosa, G.~Varoquaux, A.~Gramfort, V.~Michel, B.~Thirion, O.~Grisel,
  M.~Blondel, P.~Prettenhofer, R.~Weiss, V.~Dubourg, J.~Vanderplas, A.~Passos,
  D.~Cournapeau, M.~Brucher, M.~Perrot, and E.~Duchesnay.
\newblock Scikit-learn: Machine learning in {P}ython.
\newblock {\em Journal of Machine Learning Research}, 12:2825--2830, 2011.

\end{thebibliography}
\end{document}